\documentclass{emulateapj}
\usepackage{amssymb}
\usepackage{amsmath}
\usepackage{epsfig}
\usepackage{amsthm}
\usepackage{rotating}

\def\u{\mathbf{u}}
\def\k{\mathbf{k}}
\def\kperp{k_{\bot}}
\def\kpar{k_{\|}}

\begin{document}

\title{Foreground Contamination in Interferometric Measurements of the Redshifted 21 cm Power Spectrum}

\author{Judd D. Bowman\altaffilmark{1,2}, Miguel F. Morales\altaffilmark{3}, and Jacqueline N. Hewitt\altaffilmark{4}}

\altaffiltext{1}{California Institute of Technology, Pasadena, CA 91125, USA}
\altaffiltext{2}{Hubble Fellow}
\altaffiltext{3}{University of Washington, Seattle, WA 98195, USA}
\altaffiltext{4}{Massachusetts Institute of Technology, Kavli Institute for Astrophysics and Space Research, Cambridge, Massachusetts, USA}

\email{jdbowman@caltech.edu}

\begin{abstract}

Subtraction of astrophysical foreground contamination from ``dirty'' sky maps produced by simulated measurements of the Murchison Widefield Array (MWA) has been performed by fitting a $3^\text{rd}$-order polynomial along the spectral dimension of each pixel in the data cubes.  The simulations are the first to include the unavoidable instrumental effects of the frequency-dependent primary antenna beams and synthesized array beams.  They recover the one-dimensional spherically-binned input redshifted 21~cm power spectrum to within $\sim1$\% over the scales probed most sensitively by the MWA ($0.01 \lesssim k \lesssim 1$ Mpc$^{-1}$) and demonstrate that realistic instrumental effects will not mask the EoR signal.  We find that the weighting function used to produce the dirty sky maps from the gridded visibility measurements is important to the success of the technique.  Uniform weighting of the visibility measurements produces the best results, whereas natural weighting significantly worsens the foreground subtraction by coupling structure in the density of the visibility measurements to spectral structure in the dirty sky map data cube.  The extremely dense $uv$-coverage of the MWA was found to be advantageous for this technique and produced very good results on scales corresponding to $|u| \lesssim 500\lambda$ in the $uv$-plane without any selective editing of the $uv$-coverage.

\end{abstract}


\section{Introduction}

Astrophysical foreground contaminants are five orders of magnitude brighter than the $\sim10$ mK
redshifted 21~cm emission expected from neutral hydrogen in the intergalactic medium (IGM) during the epoch of reionization (EoR). These foregrounds will severely complicate planned experiments and the interpretation of their results.  Galactic synchrotron radiation dominates the sky at radio frequencies near 150 MHz ($z \approx 8$), accounting for $\gtrsim70\%$ of the 200 to 10,000~K total brightness temperature \citep{1999A&A...345..380S}. Extragalactic continuum point sources are also especially strong and numerous, presenting a sea of confused point sources and  comprising the bulk of the remaining $\sim30\%$ of the sky brightness temperature. Galactic radio recombination lines (RRLs) and free-free emission from electrons in both the Galaxy and the IGM will additionally complicate the planned measurements.

Initial analyses indicated that these foregrounds were an insurmountable obstacle \citep{2002ApJ...564..576D, 2003MNRAS.346..871O} because their angular variance dominates the expected fluctuations in the redshifted 21~cm background, but subsequent studies have suggested that multi-frequency observations and the application of appropriate statistical techniques should provide methods to separate the foregrounds from the redshifted 21~cm signal by exploiting the large coherence of the foregrounds with frequency \citep{2004MNRAS.355.1053D, 2004ApJ...608..622Z, 2004ApJ...615....7M, 2004NewAR..48.1039F, 2004ApJ...608..611G, 2005ApJ...625..575S, 2006ApJ...650..529W, 2006ApJ...653..815M, 2008MNRAS.391..383G}. However, these studies have relied on a number simplifying assumptions, and in particular have neglected complications due to the frequency-dependent instrumental response. The point spread function and instrumental field-of-view are both frequency dependent, and can mix the angular structure that concerned \citet{2002ApJ...564..576D} and \citet{2003MNRAS.346..871O} into the frequency direction, masking the EoR signal. This transfer of power from the angular to frequency dimensions has been dubbed ``mode-mixing'', and is a significant concern for redshifted 21~cm measurements.

In this paper we explicitly model the mode-mixing effect for confusion level contaminants and explore ways of minimizing this power transfer into the frequency domain. With these new techniques we show that, for the Murchison Widefield Array (MWA), contamination due to mode-mixing can be reduced well below that of the expected 21~cm signal strength during the reionization epoch for a reasonable foreground model.  We begin in $\S$2 by defining our instrument and sky models and establishing the mathematical foundation of the subtraction technique.  In $\S$3 we analyze the ability of the method to produce a model of the confusion-level foreground contamination and, in $\S$4, we show that the input 21~cm power spectrum in our simulations can be recovered to within $\sim1$\% (excluding thermal noise uncertainty) over most scales in the planned MWA measurements.  We conclude in $\S$5 with a brief discussion and an analytic approximation to provide context to the results of the simulation.

A detailed study of the dependence of this foreground subtraction technique on the properties of the instrument and astrophysical sky has been conducted in parallel by \cite{Adrian2008} for cases building on our fiducial MWA measurement framework.  In addition, \citet{2008MNRAS.389.1319J} have recently developed a sophisticated astrophysical foreground model and applied it to analysis of subtraction techniques in the context of the expected antenna configuration of the Low Frequency Array (LOFAR).  Their results indicate that LOFAR should be able to constrain the reionization history over a wide range of redshifts by measuring the variance in image maps from individual spectral channels between 100 and 200~MHz.  Our present focus is on foreground subtraction in the context of three-dimensional power spectrum measurements including the full frequency-dependent instrumental properties of the MWA, but many of the results of all three investigations are relevant to both measurements and instruments, as well as other active redshifted 21~cm experiments including the Precision Array to Probe the Epoch of Reionization (PAPER), the reionization project with the Giant Metre-wave Radio Telescope (GMRT), and the future Square Kilometer Array (SKA).

\section{Statement of the Problem}

Galactic synchrotron radiation, extragalactic continuum sources, and free-free emission dominate brightness temperature maps of the low-frequency radio sky and yield diffuse angular structure that is several orders of magnitude more intense than the expected redshifted 21~cm background from the reionization epoch and Dark Ages. The fundamental problem faced by all redshifted 21~cm experiments, therefore, is how to effectively isolate the desired signal at high precision from these nuisance foregrounds.  The strategies discussed in the literature for removing the foreground contamination all exploit a single, common spectral property that distinguishes the foregrounds from the expected signal. Along any one line of sight, each of the foreground components has a slowly varying power-law-like spectrum. This results in a large spectral coherence scale and is in contrast to the redshifted 21~cm signal from the reionization epoch, which fluctuates relatively rapidly in all three spatial dimensions, and thus has a short coherence scale, both in frequency and angle \citep{2004ApJ...615....7M}. In general, the spatial coherence length of the reionization signal is of order 10~Mpc, which translates to sub-degree scale fluctuations on the plane of the sky and sub-MHz fluctuations in frequency. Thus, although the angular fluctuations in the foreground intensity span the scales of interest for the 21~cm signal, the frequency fluctuations are on much larger scales than the desired signal.

In the image domain, the ``per pixel'' technique of subtracting a smooth spectral component from each pixel in a data cube should be able to effectively remove the foreground contribution \citep{2004NewAR..48.1039F, 2005ApJ...625..575S, 2006ApJ...650..529W, 2006ApJ...653..815M} and only sacrifice a few of the largest modes of the redshifted 21~cm fluctuations.  A closely related approach is to produce internal linear combinations (ILCs) of maps from different spectral channels (e.g. difference maps) in analogy with the removal of Galactic emission from CMB measurements \citep{2004ApJ...608..611G, 2004MNRAS.355.1053D}.  Unlike in CMB studies, however, template fitting of specific foreground components is not a viable approach for redshifted 21~cm foreground subtraction due to the extremely high level of accuracy needed and the scarcity of suitable templates at the target frequencies.  In the Fourier domain, one can also fit low order polynomials or derive ILCs to aid in removing foreground emission, and this has certain advantages for instruments with sparse $u,v$ coverage (Zaldarriaga et al. 2004; Ue-li Pen personal communication). In all of these approaches and the additional statistical techniques discussed in \citet{2006ApJ...648..767M}, the angular 21~cm fluctuations are effectively ignored in favor of the line-of-sight fluctuations.

\begin{figure*}
\centering
\includegraphics[width = 6 in]{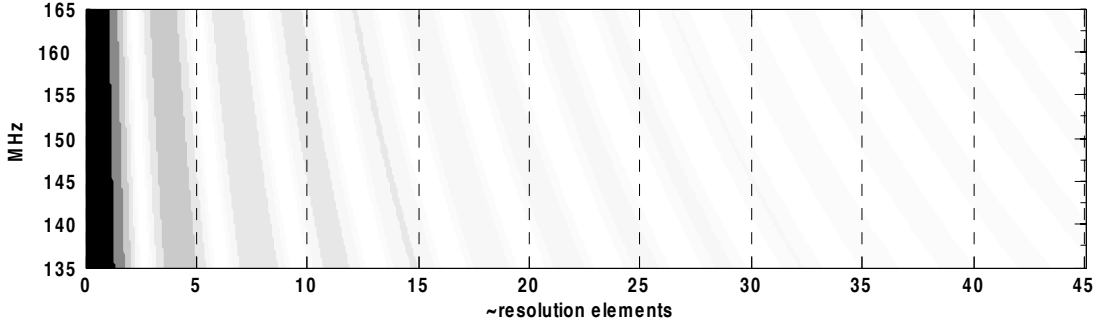}
\caption{Cartoon of a frequency vs. position plot to illustrate the origin of mode-mixing from a frequency dependent point spread function (PSF). At the origin of the horizontal axis a source has been mis-subtracted, and the residual flux ripples across the image due to the array's point spread function. Since the PSF scales with frequency, the positions of the radial intensity peaks change with frequency resulting in diagonal frequency bands. At lines of sight away from the mis-subtracted source, this leads to frequency-dependent contamination, as indicated by the contribution along the vertical dashed lines. The frequency-dependent PSF is inherent in the measurement, and mixes a pure spatial component (mis-subtracted sources) into the frequency direction. In addition, as the frequency slope of the contamination depends on the distance to the residual sources, a wide range of line-of-sight $k$-modes are contaminated.}
\label{ModeMixingFigure}
\end{figure*}

There is a significant obstacle to implementing the techniques listed above in real-world applications: inverting the instrumental response matrix for the MWA and other low-frequency arrays will not be possible.  Thus, the effects of their frequency-dependent point spread functions (PSFs) and fields-of-view (FOVs) will remain in the final data products and mix the strong angular foreground fluctuations into the frequency domain.  This instrumental effect can produce sufficient spectral structure in the derived brightness temperature maps to destroy the large spectral-coherence of the foregrounds and permanently mask the 21~cm signal. Even a modest observational bandwidth of 30 MHz at 150 MHz ($z \approx 8$) represents a fractional bandwidth of 20\%. Thus, the array appears 20\% larger in the $uv$-plane at the highest frequencies than the lowest, causing the PSF in uncorrected maps to change in angular size by 20\% from the one end of the frequency band to the other.  The FOV is similarly 20\% larger at the bottom of the band than the top.\footnote{One can design antennas with a frequency-dependent collecting area to counteract the effect of the aperture becoming smaller in wavelengths and produce a near constant FOV. However, for a number of technical reasons, none of the upcoming arrays use this kind of detector element.} Consequently, ripples and peaks introduced into the image by the instrumental PSF scale with frequency and appear in different locations at different frequencies. This is the mechanism that couples angular fluctuations into the frequency domain, and is shown graphically in Figure \ref{ModeMixingFigure}. Similarly, the varying FOV significantly changes the input sky signal received at different frequencies. This is particularly important for Fourier domain subtraction techniques, where a 20\% change in FOV implies that $\sim$40\% of the signal is not in common for a visibility measured at the same point in the $uv$-plane, but at opposite ends of the band.

In order to study the effect of mode-mixing, we need to carefully simulate the instrument, foreground contamination, and analysis pipeline. In the following sections we describe the simulation parameters used for this study.

\subsection{Epoch of Reionization Observations with the Murchison Widefield Array (MWA)}

    \begin{deluxetable}{lr}
    \tablecaption
    {
        MWA Design Specifications
        \label{tab_detector}
    }

    \tablehead{\colhead{Parameter} & \colhead{Value}}

    \startdata
    Array layout, $\rho(r)$ (m$^{-2})$     & $\sim r^{-2}$\\
    Array diameter, D (m)                       & 1500 \\
    Array latitude ($^\circ$)                   & -27 \\
    Bandwidth, B (MHz)                          & 32 \\
    Spectral resolution (kHz)                   & 40 \\
    Number of antennas, N                       & 500 \\
    Angular resolution, $\Theta_B$ ($^\circ$)   & 0.073 \\
    Antenna collecting area, dA (m$^2$)         & 14.4 \\
    Antenna response scale, $\Theta_P$ ($^\circ$) & 31 \\
    Central frequency, $\nu_0$ (MHz)            & 158 \\
    System temperature, $T_{sys}$ (K)           & 440 \\
    \enddata

    \tablecomments
    {
        Parameters are listed with their fiducial values at $\nu=158$~MHz, corresponding to $z=8$ for redshifted 21~cm measurements.
    }

    \end{deluxetable}

The fiducial MWA design is described in \citet[Their Section 2]{2006ApJ...638...20B}. The array design consists of $N=500$ antennas distributed within a $D=1500$ m diameter circle. The density of antennas as a function of radius is taken to go as $\sim r^{-2}$, but capped at a maximum density of one antenna per 36~m$^2$. The antenna response is approximated by
\begin{equation}
    \label{eqn_window}
    \begin{array}{cc}
        W(\theta) = \cos^2 \left ( \frac{\pi}{2} \theta / \Theta_P \right ), & \quad \mbox{ $\theta < \Theta_P$} \\
    \end{array}
\end{equation}
where $\Theta_P$ is proportional to wavelength and is $31^\circ$ at $\nu=158$~MHz. The angular resolution of the array is given by $\lambda/D$ and the total collecting area by $N~dA$, where $dA$ is the collecting area of each antenna and scales like $dA = 16 ( \lambda^2/4 )$ for $\lambda < 2.1$~m and is capped for longer wavelengths. Finally, the full instantaneous bandwidth of the instrument is $B=32$~MHz and the spectral resolution is $10$~kHz, although the EOR observations will be binned to $\geq40$~kHz resolution to reduce the data volume. All of the fiducial properties are summarized in Table \ref{tab_detector}.  For the analysis in this paper, we define the observation to be of a single field with 360~hours integration during the most favorable circumstances. Additionally, we set the frequency coverage to $142<\nu<174$ MHz, which spans $9 > z > 7.1$.

The foreground subtraction strategy planned for redshifted 21~cm measurements with the MWA is a multi-stage process consisting of three primary components:

\textit{1 - Bright source subtraction.}  The first step is to subtract individual bright sources using a full deconvolution approach that incorporates the position, frequency, and time dependent antenna calibrations to remove the sources to high precision.  The MWA has been designed to aid in the mitigation of extremely intense extragalactic and Galactic continuum sources.  It is common practice in radio astronomy to ``peel'' away bright sources in the field of view to improve  the overall dynamic range in derived maps of the sky brightness and, thus, to increase the sensitivity to fainter signals. This process must be applied to the bright point sources in measurements by the MWA to a high degree of precision (better than 1 part in $10^5$) in order to reveal the fluctuations in the redshifted 21~cm background. For typical interferometers, the sparse coverage of visibility measurements in the $uv$-plane means that a substantial amount of power from these point sources is spread across the derived map of the sky due to the side lobes of the point spread function of the synthesized beam.  This poses a considerable deconvolution problem and makes it difficult to subtract them from the measurements since the locations and intensities of the point sources must be determined very accurately before their contributions to the individual visibility measurements can be removed to high  precision. The large number of antennas used for the MWA reduces the side lobes of the synthesized beam and provides high angular dynamic range in uncorrected maps of the sky. This is advantageous for isolating and subtracting the power from extremely intense extragalactic continuum sources and alleviates the severity of the bright source
contamination.

\textit{2 - Diffuse spectral subtraction.} The next step is to remove diffuse emission from the Galaxy and faint confusion-level sources by fitting a polynomial to the spectrum along each line of sight in the observed sky maps in order to subtract the spectrally smooth foreground component.  To aid this process, the angular size of the primary antenna beam (field of view) for the MWA has been matched to fit within the relatively cold regions that exist in the Galactic synchrotron emission at high Galactic latitudes. Figure~\ref{f_foregrounds_synchrotron} illustrates the typical brightness temperature of the synchrotron  foreground at 150~MHz, as well as the primary field of view for an MWA antenna tile that is targeting a cold part of the sky.

\textit{3 - Statistical template subtraction.} The final step in the foreground subtraction process is envisioned to take place as part of the 21~cm power spectrum analysis by fitting templates of the residual statistical structures of foregrounds that may remain even after the first-order subtraction of their emission from sky maps.

In the analysis that follows, we focus only on the second step: subtracting faint sources and diffuse emission that cannot be identified as individual contributions in the maps and deconvolved efficiently.  We do not model the first step of bright source deconvolution, but rather assume it has been performed perfectly.  At the end of this paper, we look briefly at the implications from the present analysis for the final planned foreground subtraction step of statistical template fitting.

\subsection{The Sky Model}
\label{skyModelSec}

In this section, we specify the details of the sky model for the astrophysical foregrounds and the redshifted 21 cm signal.  We build our foreground sky model adhering closely to existing empirical findings.  The underlying physical mechanisms behind the observed properties are not addressed in the development of this model since we are interested primarily in the functional consequences of their manifestations on the sky and in the instrument (for an alternative treatment see \citet{2008MNRAS.389.1319J}).  We begin by constructing an empirically-motivated model of the astrophysical foregrounds that consists of two components: 1) discrete, faint extragalactic continuum sources and 2) diffuse Galactic synchrotron emission.  Together, these are expected to account for approximately 98\% of the total intensity in the radio spectrum below 200~MHz \citep{1999A&A...345..380S, 1967MNRAS.136..219B}.  We exclude free-free emissions as a separate component in our analysis since they have power-law spectra similar to the other components \citep{1996ApJ...460....1K} and they are easily subsumed by the uncertainty in the discrete continuum source contribution, although the free-free emissions have been shown to correlate with dust clouds at high Galactic latitudes, and thus have different angular structure  than the discrete source population. We have also chosen to ignore supernova relics, radio clusters, and RRLs in this analysis. The first two categories of sources do not differ substantially from the power-law-like components we  include in the model, and the frequencies at which Galactic RRLs occur are known and can be excised from observations if they are determined to be a significant contribution. Diffuse RRLs have never been observed at high Galactic latitudes, but they are expected to be narrower than the 40~kHz frequency channel size of the MWA and occur approximately every few MHz, and therefore in only a few percent of the spectral channels  [\citet{1995ApJ...454..125E}, \citet{1975Prama...5....1S} and \citet{1982ApJ...260..317W}, are good starting points for additional details about the observed properties and theoretical treatment of RRLs, respectively]. Thus, the cost of excising the RRLs is a minor complication to the window function, but one that is expected to be no more severe than that caused by excising radio-frequency interference (RFI).

\subsubsection{Diffuse Galactic Synchrotron Radiation}

\begin{figure*}
\centering
\includegraphics[width=20pc, angle=-90]{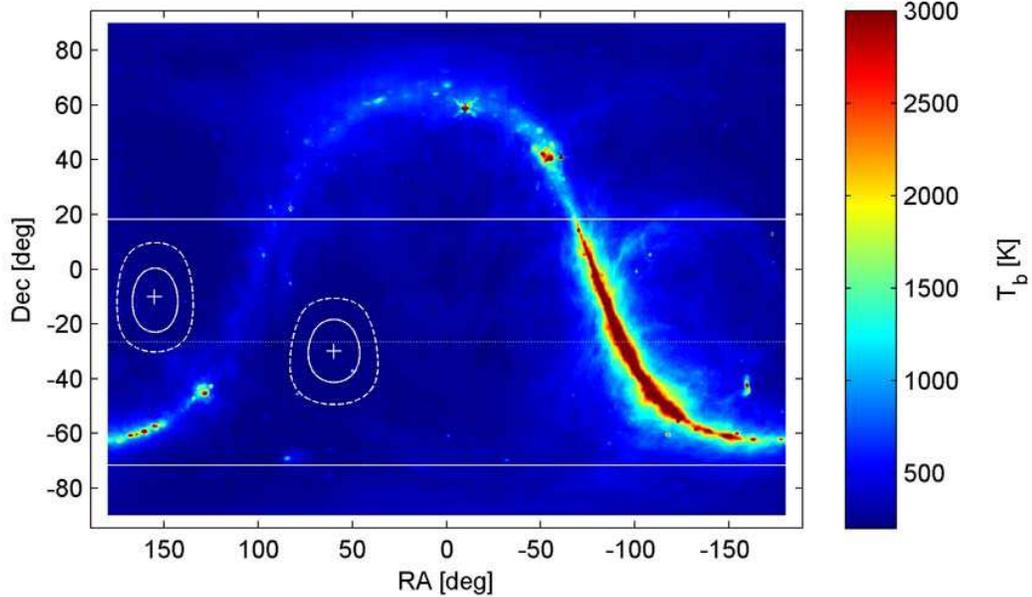}
\caption[Brightness temperature of the Galactic synchrotron foreground]{ \label{f_foregrounds_synchrotron} Brightness temperature of the Galactic synchrotron foreground at $\nu=150$~MHz. The all-sky measurements of \citet{1982A&AS...47....1H} were scaled from 408~MHz using a constant spectral index of $\beta=2.6$ to produce this map. The declination of the zenith at the MWA latitude is shown (horizontal dotted line), as is the declination range within zenith angle of 45$^\circ$ (horizontal solid lines). The MWA will have access to cold regions ($\sim250$~K) of the sky both north and south of the Galactic plane.  The primary beam of the MWA antenna tiles is shown centered at the two planned target fields, marked by plus symbols at 60 R.A. by -30 declination and 115 R.A. by 10 declination.  The concentric circles indicate the 50\% (solid), and 10\% (dashed) power response relative to the peak response. }
\end{figure*}

For the diffuse Galactic synchrotron foreground, there are a number of observations on which to base the model \citep{1967MNRAS.136..219B, 1970AuJPA..16....1L, 1982A&AS...47....1H, 1988A&AS...74....7R, 1997A&AS..124..315A, 1999A&AS..137....7R}.  The spectrum of the Galactic synchrotron emission has been found to be a nearly featureless power-law with modest variations in the spectral index as a function of direction and frequency.  At the frequencies of interest, the spectral index is given by $T\sim\nu^{-\beta}$, and has a typical value of $\beta\approx2.5$ \citep{2008AJ....136..641R}. In general, it is steeper at high Galactic latitudes than toward the Galactic plane.  The spectral index also steepens as a function of frequency to a maximum of $\beta\approx2.8$ by $\nu\approx1$~GHz, but is generally constant below about 200~MHz.  The variation of the spectral index across the sky has a standard deviation of order $\sigma_\beta\approx0.1$ on degree scales \citep{1967MNRAS.136..219B, 1988A&AS...74....7R, 1998ApJ...505..473P, 1999A&AS..137....7R} and appears to be weakly correlated to the angular structure \citep{1986A&AS...63..205R, 1987MNRAS.225..307L, 2003A&A...410..847P}.  Angular structure in the diffuse Galactic emission has been shown to be well-described by a power-law spectrum in Fourier space over a large range of scales.  The angular power-law index is specified according to $C_{\ell}\sim\ell^{-\alpha}$, where typically $2.4\lesssim \alpha \lesssim2.9$ \citep{1998ApJ...505..473P, 2001A&A...371..708G, 2002A&A...387...82G, 2008A&A...479..641L}, and the spherical harmonic multipole is related to the $uv$-plane baseline length by $\ell \approx 2 \pi u$.

\begin{figure}
\centering
\includegraphics[width=18pc,angle=-90]{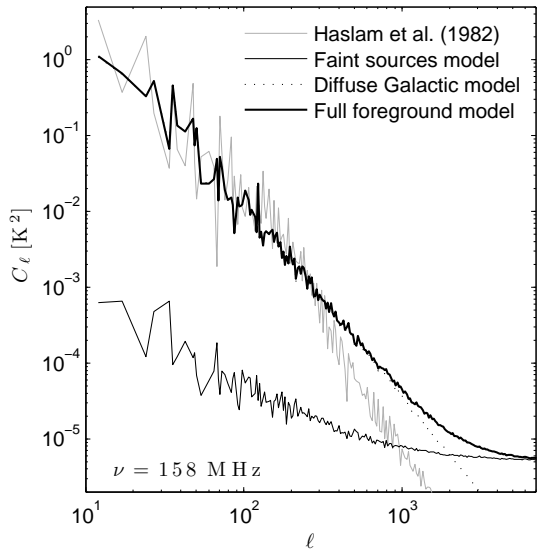}
\caption[Angular power spectra of foreground components]{
\label{f_foregrounds_aps} Angular power spectra of the two model foreground components along with the power spectrum derived from the \citet{1982A&AS...47....1H} map (scaled to 150~MHz) over the region spanned by the MWA primary target field.  The diffuse Galactic synchrotron emission dominates the foreground model at large scales (low $\ell$), where its angular fluctuation amplitude and power-law index have been matched to the \citet{1982A&AS...47....1H} map.  For small, sub-degree scales ($\ell\gtrsim200$), the power in the \citet{1982A&AS...47....1H} map drops off rapidly due to its limited angular resolution, whereas the Galactic component of the foreground model continues its pure $\alpha=2.55$ power-law profile.  At very small scales ($\ell\gtrsim1000$), the angular power of the extragalactic continuum source model becomes comparable to the diffuse Galactic power since the continuum source fluctuations have a Poisson-noise floor. }
\end{figure}

We construct our model of the diffuse Galactic synchrotron radiation in two steps.  First, we produce a Gaussian random field with angular power-law index of $\alpha=2.55$.  The amplitude of this field is normalized to both the mean and angular fluctuation power in the region of the all-sky map of \citet{1982A&AS...47....1H} centered on $R.A.=60^\circ$ and $Dec.=-30^\circ$. This region has been chosen as the primary target field for the MWA due to its relatively low brightness temperature and its location roughly opposite the Galactic center. In order to produce appropriate amplitude normalizations, the temperatures in the \citet{1982A&AS...47....1H} map are scaled from 408~MHz to 150~MHz using a constant spectral index of $\beta=2.6$. The results of this scaling are shown in Figure~\ref{f_foregrounds_synchrotron}.  Since the \citet{1982A&AS...47....1H} map is lacking angular information on scales smaller than about 1$^{\circ}$, only multipoles below $\ell<200$ were used to normalize the angular fluctuation power in our model.  Figure~\ref{f_foregrounds_aps} illustrates the angular power spectrum of our model at 150~MHz, as well as the power spectrum derived from the \citet{1982A&AS...47....1H} map.  To complete our synchrotron model, the second step is to assign a spectrum for each line-of-sight in the simulated map.  We specified each spectrum as a power law with spectral index drawn from a second Gaussian random field.  The spectral index field has $\bar{\beta}=2.6$ and angular structure given by the sum of two terms: 1) a white noise contribution with $\sigma_{\beta}=0.1$ and 2) the simulated brightness temperature map at 150~MHz renormalized such that it yields an additional variance in the spectral index map of $\sigma_{\beta}=0.05$.  Thus, the spectral index map is weakly correlated with the brightness temperature map.  The level of correlation was set by eye from inspection of derived brightness temperature and spectral index maps including, in particular, the maps of \citet{2003A&A...410..847P}, although we note that those results were found at higher frequencies.

\subsubsection{Extragalactic Continuum Sources}
\label{sec_foregrounds_pointsources}

Normal galaxies, radio galaxies, and active galactic nuclei, form the majority of the extragalactic continuum sources \citep{2005ApJ...625..575S}.  Their contribution to the brightness temperature has been found to be of order $30$ to $70$~K at 178~MHz by \citet{1967MNRAS.136..219B} by fitting an isotropic background component to measurements of spatial variations in the spectral index of the diffuse emission between 18 and 404~MHz. In the same analysis, it was found that the integrated brightness temperature due to extragalactic sources has a characteristic spectral index of $\beta=2.7$, although with $\sim10$\% uncertainty, and therefore is somewhat steeper than the Galactic synchrotron spectral index. Individual sources have been observed to have significantly more variation in their spectral indices, however, spanning $-2 \lesssim \beta \lesssim 3$.  The inverted cases ($\beta\approx-2$) are due to
sources with strong synchrotron self absorption, such that the spectrum has already flattened at the frequencies of interest (as opposed to flattening below $\sim10$~MHz, as would be more typical).
In general, the spectral index of discrete extragalactic sources is related to the intensity of the source.  It is steeper for brighter sources (with maximum $\beta\approx3$) than for fainter sources, where $\beta\approx2.5$ is typical for sources with fluxes less than $S<0.1$~Jy \citep{1969ApJ...157....1K, 2003A&A...404...57Z, 2004ApJS..150..417C}.

A number of surveys of radio sources have been performed at frequencies relevant to the MWA (see \citet[their Fig. 1]{2004ApJS..150..417C}). They include a survey at 74~MHz by \citet{2004ApJS..150..417C}, the 7C \citep{1990MNRAS.246..110M} and 6C surveys \citep{1988MNRAS.234..919H} at 151~MHz, the 3CR survey \citep{1983MNRAS.204..151L} of bright ($>10$~Jy) sources at 178~MHz, WENSS \citep{1997A&AS..124..259R} at 327~MHz, and the 5C5 survey \citep{1975MNRAS.171..475P} at 408~MHz, as well as the FIRST survey \citep{1997ApJ...475..479W} and NVSS \citep{1998AJ....115.1693C} at 1.4~GHz. Analysis of the catalogs of radio sources derived from these and other surveys have yielded a solid understanding of the statistical properties of radio sources \citep{vla-sci-146, 1985ApJ...289..494W, 1991PhDT.......241W, 1991AJ....102.1258F, 1993ApJ...405..498W}.  The brightest sources in the catalogs are relatively nearby objects and, thus, have differential source counts given by $N(S) \sim S^{-2.5}$, as would be expected for a population of objects distributed evenly in a Euclidean space. Weaker sources are generally more distant and the observed differential counts fall off more gradually, as $N(S)\sim S^{-1.8}$, due to evolution of the population and redshift effects. At extremely faint flux levels (in the $\mu$Jy regime), a new population of local starburst galaxies becomes visible and the differential counts again become steeper, with $N(S)\sim S^{-2.2}$. Differential source counts cannot asymptote to this power-law as $S\rightarrow0$, however, since it is steeper than $N(S)\sim S^{-2}$ and an infinite radio flux would result. The differential counts must eventually flatten again with decreasing flux.

For our extragalactic continuum source model, we first produce a simulated catalog of radio sources and then draw from that catalog to sum over individual sources within each pixel of a simulated sky map. We use the expression for differential source counts given by \citet{2002astro.ph..9569S},
\begin{equation}
N(S) = 100 ~  S_{\mu \text{Jy}}^{-2.2} ~ \nu_{\text{GHz}}^{-0.8} ~
\text{arcmin}^{-2} ~ \mu\text{Jy}^{-1} \label{eqn_source_counts}
\end{equation}
where $S_{\mu \text{Jy}}$ is the flux measured in $\mu$Jy and $\nu_\text{GHz}$ is the frequency measured in GHz. Based on this equation, we estimate that it should be possible to identify individual sources in sky maps produced by the MWA down to a flux limit of approximately 10~mJy before the sources become confused (in other words, there will be more than approximately one source per pixel below 10~mJy and, hence, individual sources will not be identifiable).  Since the integrated flux due to faint radio sources in Equation~\ref{eqn_source_counts} converges around $10$~$\mu$Jy, and since we want to limit the number of sources in our simulated catalog, we only model sources between $10$~$\mu$Jy~$<S<10$~mJy.  For each source in our simulated catalog, we randomly assign a flux such that Equation~\ref{eqn_source_counts} is satisfied.  We also randomly assign a spectral index for each source from a distribution of spectral indices that is the sum of two Gaussian distributions.  One distribution has $\bar{\beta}=2.7$ and $\sigma_\beta=0.1$, and the second distribution has $\bar{\beta}=0$ and $\sigma_\beta=0.6$ in
order to account for the large variation of spectral indices in extragalactic continuum sources.  Only 10\% of the sources have their spectral indices drawn from the second distribution.

In order to produce a realistic sky map from our simulated source catalog, the angular correlation of radio sources must be taken into account. \citet{2002MNRAS.329L..37B} analyzed the 1.4~GHz NRAO VLA Sky Survey (NVSS) and found evidence for weak clustering of the continuum point sources.  The angular correlation function that they derived is
\begin{equation}
\label{eqn_angular_corr} w(\theta) = 10^{-3} ~ \theta^{-0.8},
\end{equation}
for $\theta$ measured in degrees. The magnitude of this clustering is much lower than for the galaxy correlation functions determined optically, and is the result of projection effects.  Deep radio catalogs contain sources spanning a large range of redshifts since their continuum spectra do not allow for grouping by photometric redshift measurements.  This causes the angular correlation function of radio continuum sources to dilute since unrelated volumes of the Universe are superimposed in the observations.  We follow the method of \citet{2005ApJ...621....1G} to make a realization of the source counts in pixels that includes both the Poisson variance and the proper angular clustering.  The large scale structure introduced by the angular clustering adds an additional variance to the distribution of point sources and the resulting brightness temperature in the sky map.  Using this source count map, we populate each line-of-sight in the extragalactic continuum source foreground model by drawing the appropriate number of sources from our simulated catalog and summing their contributions. Figure~\ref{f_foregrounds_aps} shows the angular power spectrum resulting from this process at 150~MHz.

\subsubsection{Redshifted 21~cm Signal}
\label{sec_foregrounds_signal}

Since the purpose of this work is to study the effects of a realistic interferometer on foreground subtraction, we employ a simple model for the redshifted 21 cm signal.  Averaging over velocity field distortions, the brightness temperature of diffuse redshifted 21~cm emission can be described by,
\begin{equation}
\begin{array}{rl}
\label{eqn_intro_temp} \delta T_{21}(\vec{\theta}, z) \approx~&
23~(1+\delta)~x_{HI} \left ( 1 - \frac{T_{CMB}}{T_S} \right ) \\
& \times \left ( \frac{\Omega_b~h^2}{0.02} \right ) \left [ \left (
\frac{0.15}{\Omega_m~h^2} \right ) \left ( \frac{1+z}{10} \right )
\right ]^{1/2} \mbox{mK},
\end{array}
\end{equation}
where $\delta$ is the matter density perturbation field, $x_{HI}$ is the neutral fraction of hydrogen, $T_{CMB}=2.725$~K is the CMB temperature, and $T_S$ is the spin temperature describing the relative population of the 21~cm hyperfine states.  For our simple model, we assume that the IGM remains fully neutral at our target redshift of $z=8$ and that the spin temperature of neutral hydrogen is $T_S \gg T_{CMB}$.  The only fluctuations in the 21~cm brightness temperature, therefore, are due to matter density perturbations, which we model as a Gaussian random field with a power spectrum given by the output of CMBFAST \citep{1996ApJ...469..437S} and scaled to the target redshift according to the linear growth factor.

\subsection{Instrumental Simulation}
\label{sec_foregrounds_method}

We start with our model of the frequency-dependent sky brightness $I(\{\vec{\theta}, \nu\})$, as detailed in \S \ref{skyModelSec}. For the remainder of this section we will drop the explicit frequency dependence of all the terms to make the notation clearer. The result of observing the true sky with the MWA is a vector, $\mathbf{m}$, of complex visibility measurements.  This process can be expressed as
\begin{equation}
\label{eqn_measurement} \mathbf{m}(v) = \mathbf{M}(\vec{\theta},\vec{\theta})
~I(\vec{\theta}) + n(v),
\end{equation}
where $v$ indexes the individual measurements, $\mathbf{M}$ is the instrumental response matrix that maps the true sky to the resulting measurements, and $n$ is a vector containing the system noise added to each measurement.

In observations with the MWA the visibility data must be calibrated and compressed so that long integrations can be archived for later analysis, and this process is detailed in \citet{2008arXiv0810.5107M, 2008arXiv0807.1912M}.  Assuming perfect calibration, the resulting images and Fourier representations can be simulated as
\begin{eqnarray}
I'(\vec{\theta}) = \mathbf{B}( \vec{\theta},\vec{\theta})\ I(\vec{\theta}) + n(\vec{\theta}) \label{dirtyImgEq}\\
I'(\u) = \mathbf{B}( \u,\u)\ I(\u) + n(\u). \label{dirtyFEq}
\end{eqnarray}
Equations \ref{dirtyImgEq} and \ref{dirtyFEq} are Fourier transforms of one another and the operator $\mathbf{B}$ represents the frequency dependent array beam and antenna field of view and $I$ is the true sky in either angular or transformed coordinates. $I'$ in the first line is the standard interferometric dirty map, and the gridded $uv$-plane in the second line. Depending on the application we will alternate between these two forms of the equation. The first line is useful when fitting spectra to individual image pixels, but has the disadvantage of a highly covariant noise matrix $N = n^{T}(\vec{\theta})n(\vec{\theta})$. The second line has (to good approximation) a diagonal noise matrix that is simply related to the density of visibility measurements in the $uv$-plane.

\begin{figure*}
\centering
\includegraphics[height=42pc, angle=-90]{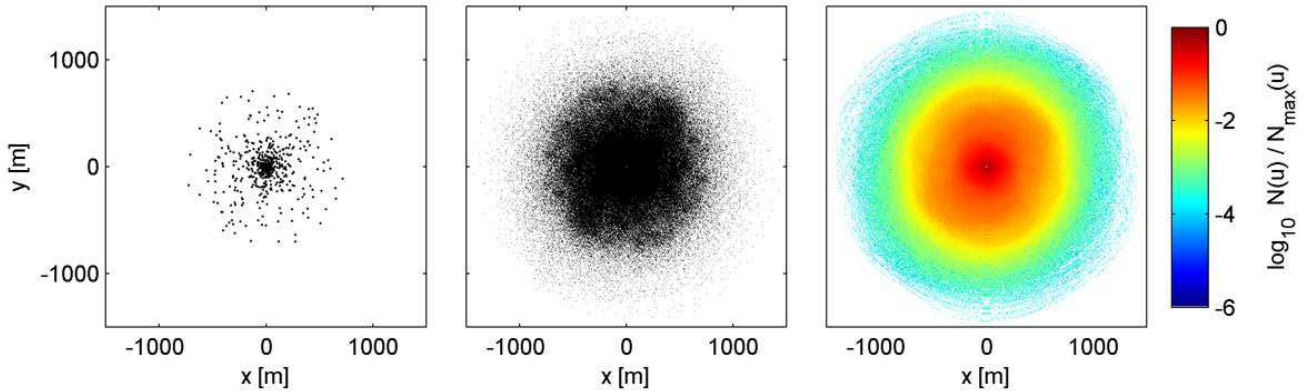}
\caption[Realization of the MWA antenna layout and corresponding
baseline distribution]{ \label{f_foregrounds_antdist} Realization of
the MWA antenna layout (left) and corresponding distribution of baselines (middle).
The 500~antennas are nearly coplanar and distributed over a 1500~m diameter
region with a density that falls off as $\sim r^{-2}$. The baselines in the
$uv$-plane are formed by the pairwise combination of all the antennas for
an instantaneous observation of a target at the zenith.  The right panel illustrates the relative density
of visibility measurements after earth-rotation synthesis has been
performed by tracking a target field at $R.A.=-60^\circ$ and
$Dec.=-30^\circ$ for six hours as it transits. }
\end{figure*}

Relating these expressions to the polynomial-based foreground subtraction investigations of \citet{2004NewAR..48.1039F}, \citet{2006ApJ...650..529W}, and \citet{2008MNRAS.391..383G}, we find that these works treated simplified cases of this general approach by assuming complete coverage of the visibility measurements in the $uv$-plane, equivalent to using $\mathbf{B}=\mathbf{I}$, where $\mathbf{I}$ is the identity matrix, and by setting $n(\u)=0$ or assuming $n(\u)$ to be constant so that the noise covariance matrix, $N$, is diagonal in the image domain, even though this is not attainable for a realistic distribution of antenna elements. \citet{2008MNRAS.389.1319J} have improved on these initial approximations by including the planned coverage of visibility measurements in the $uv$-plane for LOFAR, although the frequency dependence of their $uv$-coverage is specifically removed by excising baselines lacking corresponding measurements at all frequencies and they neglect the frequency-dependence of the primary antenna beam.   Here, we extend the growing foundation of these efforts by investigating the full frequency-dependent effects of including a realistic antenna distribution and primary antenna FOV for the MWA to calculate $\mathbf{B}$ and $n$. We expand $\mathbf{B}$ into two primary constituents
\begin{equation}
\label{eqn_response} \mathbf{B} = \mathbf{B}_{\rm array} ~
\mathbf{B}_{\rm ant},
\end{equation}
where $\mathbf{B}_{\rm array}$ accounts for the synthesized array beam due to the coverage of the visibility measurements in the $uv$-plane and $\mathbf{B}_{\rm ant}$ accounts for the response of the primary beam of the antenna tiles.  For $\mathbf{B}_{\rm ant}$ we use the window function in Equation~\ref{eqn_window}, and for $\mathbf{B}_{\rm array}$ we compute the typical $uv$-coverage of an MWA observation including earth-rotation synthesis for a $90^{\circ}$ track through zenith as shown in Figure~\ref{f_foregrounds_antdist}.

For the noise contribution to the simulated observations, we take $n(v)$ to be a vector of Gaussian random variables with standard deviation \citep{2005ApJ...619..678M}
\begin{equation}
\label{eqn_foregrounds_noise} \sigma_n = \frac{2 k_B
T_{sys}}{dA~\sqrt{d\nu~\tau}},
\end{equation}
where $k_B$ is the Boltzmann constant, $T_{sys}=440$~K is a conservative estimate of the system temperature (including sky noise) of the MWA at $\nu=158$~MHz, $dA=14.4$~m$^2$ is the effective collecting area of an MWA antenna tile at the same frequency, $d\nu=40$~kHz is the spectral resolution of a single channel, and $\tau=8$~s is the accumulation
duration for each visibility measurement.  We approximate $n(\u)$ by pixelizing the $uv$ plane with cells equal in area to the inverse of the antenna FOV. The contribution of any single visibility measurement is then applied to only one grid cell. In this limit, the thermal noise is uncorrelated in the Fourier domain maps and given by, $n(\u) = \sigma_n / \sqrt{N(\u)}$, where $N(\u)$ is the number of visibility measurements contributing to each grid cell.  Using the radiometer equation with the mean filling fraction for the MWA antenna configuration gives a fiducial thermal uncertainty for a derived image map of $\sim335$~mK after 360~hours, although in practice the thermal noise is dependent on angular scale size and improves rapidly as scale size increases due to the highly condensed antenna distribution such that at degree-scales the thermal uncertainty is only $\sim5$~mK.


The final component of modelling the foreground subtraction is to actually fit and subtract the foreground model from the sky brightness map derived from the observations. There are several reasonable methods that can be used to derive the sky brightness from the archived measurements. The most ideal map to use would be a minimally biased estimate of the true sky.  This would require inverting the instrumental response matrix, $\mathbf{B}$, and the noise covariance matrix \citep{1997ApJ...480...22T}.  In practice, it will not be feasible to invert $\mathbf{B}$ and our goal is to set a worst-case limit due to the effects of the instrumental response of the MWA on subtracting foreground contamination, and a minimally biased estimate of the sky should remove many of these effects.  Thus, we turn instead to what is commonly called visibility weighting of the dirty sky map in radio astronomy. This is the map that results by simply multiplying the $uv$ representation by a position-dependent weight, or equivalently Wiener filtering. The weighted dirty sky map is given by
\begin{equation}
\label{eqn_dirty_map}I'_U(\theta) = \mathbf{F}(\theta, \u) ~
\mathbf{U}(\u, \u) ~ I'(\u),
\end{equation}
where $\mathbf{F}$ is the Fourier transform operator and $\mathbf{U}$ is a weighting function for each pixel in the Fourier map \citep{vla-sci-154, 1999ASPC..180..127B, thompson_moran_swenson}.  The smallest uncertainty in the derived sky map due to thermal noise is achieved by weighting by the inverse of the variance due to the (Gaussian) thermal noise in each pixel in the Fourier domain. This is commonly called natural weighting, and in this formalism corresponds to $\mathbf{U} = \mathbf{B}$ if the noise in each visibility is the same. Although it produces the least uncertainty, this method typically emphasizes the information contained in short baselines because short baselines are more numerous in radio interferometers than long baselines. Thus, the effective resolution of the derived sky brightness map is lower than the $\lambda/D$
expectation. Another common method, called uniform weighting, alleviates this under-resolution at the expense of introducing more uncertainty into the derived sky map. Uniform weighting, like its name suggests, is constant for each pixel in the Fourier domain. This allows the information in the long baselines to be emphasized and restores the effective resolution of the sky map to the expected level. Figure~\ref{f_foregrounds_weightprofiles} illustrates the
difference in the weighting functions.  We will test both of these weighting methods, as well as an approximation to the natural weighting function derived by fitting a double exponential to the distribution of the number of baselines per $uv$-cell.  We will see if the foreground subtraction technique is robust to any effects the weighting scheme
might introduce.

Once a dirty sky map has been generated, a low-order polynomial is fit to each pixel in the map and subtracted. The residual contamination, $r(\theta)$, after this subtraction is given by
\begin{equation}
r_{U}(\theta) = I_U(\theta) - d(\theta) \label{eqn_resdef}
\end{equation}
where $d(\theta)$ is the ``dirty'' foreground model generated from the polynomial fits.  However, regardless of the weighting used to determine the model $d(\theta)$, we want to subtract this model from the unweighted data to preserve the signal-to-noise ratio of the original signal.  Leaving the weighting function in the Fourier domain, the residual in the unweighted data is given by
\begin{eqnarray}
r'(\theta) & = & ~\mathbf{F}^{-1}(\theta,\u)\mathbf{U^{-1}}(\u, \u)~\mathbf{F}(\u, \theta)~r_{U}(\theta) \nonumber\\
      & = &  ~\mathbf{F}^{-1}(\theta,\u)\mathbf{U^{-1}}(\u, \u)~\mathbf{F}(\u, \theta) \left [ I_U(\theta)~ -~ d(\theta) \right ] \nonumber\\
      & = & ~I'(\theta)~-~\mathbf{F}^{-1}(\theta,\u)\mathbf{U^{-1}}(\u, \u)~\mathbf{F}(\u, \theta)~d(\theta)\nonumber\\
r'(\theta)      & = & ~I'(\theta)~-~d'(\theta) \label{eqn_resnat}
\end{eqnarray}
where $d'(\theta) = \mathbf{F}^{-1}(\theta,\u)\mathbf{U^{-1}}(\u, \u)~\mathbf{F}(\u, \theta) ~
d(\theta)$.  The residual, $r'$, in the Fourier domain map is the
quantity we are interested in determining, and that we hope will yield the expected redshifted 21~cm
fluctuations.

\section{Measuring the Confusion Level Foreground}
\label{sec_foregrounds_subtraction}

\begin{figure}
\centering
\includegraphics[width=22pc]{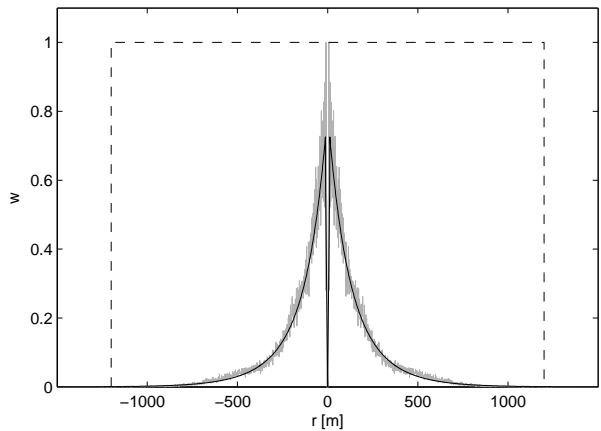}
\caption{ \label{f_foregrounds_weightprofiles} Illustration of
three different weighting functions used for forming the dirty sky
maps in Figures~\ref{f_foregrounds_sky_inst_uniform}
through~\ref{f_foregrounds_sky_inst_exponential}.  Each profile is for a
north-south cross-section through the center of the baseline
distribution after earth-rotation synthesis on the target field, and
are for uniform weighting (dash line), natural weighting (gray line) and the smoothed double exponential approximation to natural weighting (black line). The profiles are
normalized in this plot such that the maximum weight is one.  The
natural weighting profiles favor the short baselines, while the
uniform weight accentuates the long baselines.}
\end{figure}

For all the cases discussed in this section, a $3^\text{rd}$-order polynomial is used to fit the full 32~MHz of observed spectrum along each pixel of the sky map according to
\begin{equation}
d\left(\vec{\theta}, \nu \right) = a_3(\vec{\theta}) \nu^3 + a_2(\vec{\theta}) \nu^2 + a_1(\vec{\theta}) \nu + a_0(\vec{\theta})
\end{equation}
where $d$ is the model fit introduced in Equation~\ref{eqn_resdef}, and the coefficients $a_n$ are solved on a per pixel basis to minimize the least squared difference between the weighted dirty sky map, $I_U$, and the polynomial fit.  The spectral fits are performed in linear coordinates.  Fitting polynomials in logarithmic space has been considered by \citet{2006ApJ...650..529W} and others, but we find that, since the response of the interferometer acts to remove the mean value of each spectral channel and causes the value of some pixels to be less than zero, logarithmic fits pose extra complications.

\begin{figure*}
\centering
\includegraphics[width=38pc]{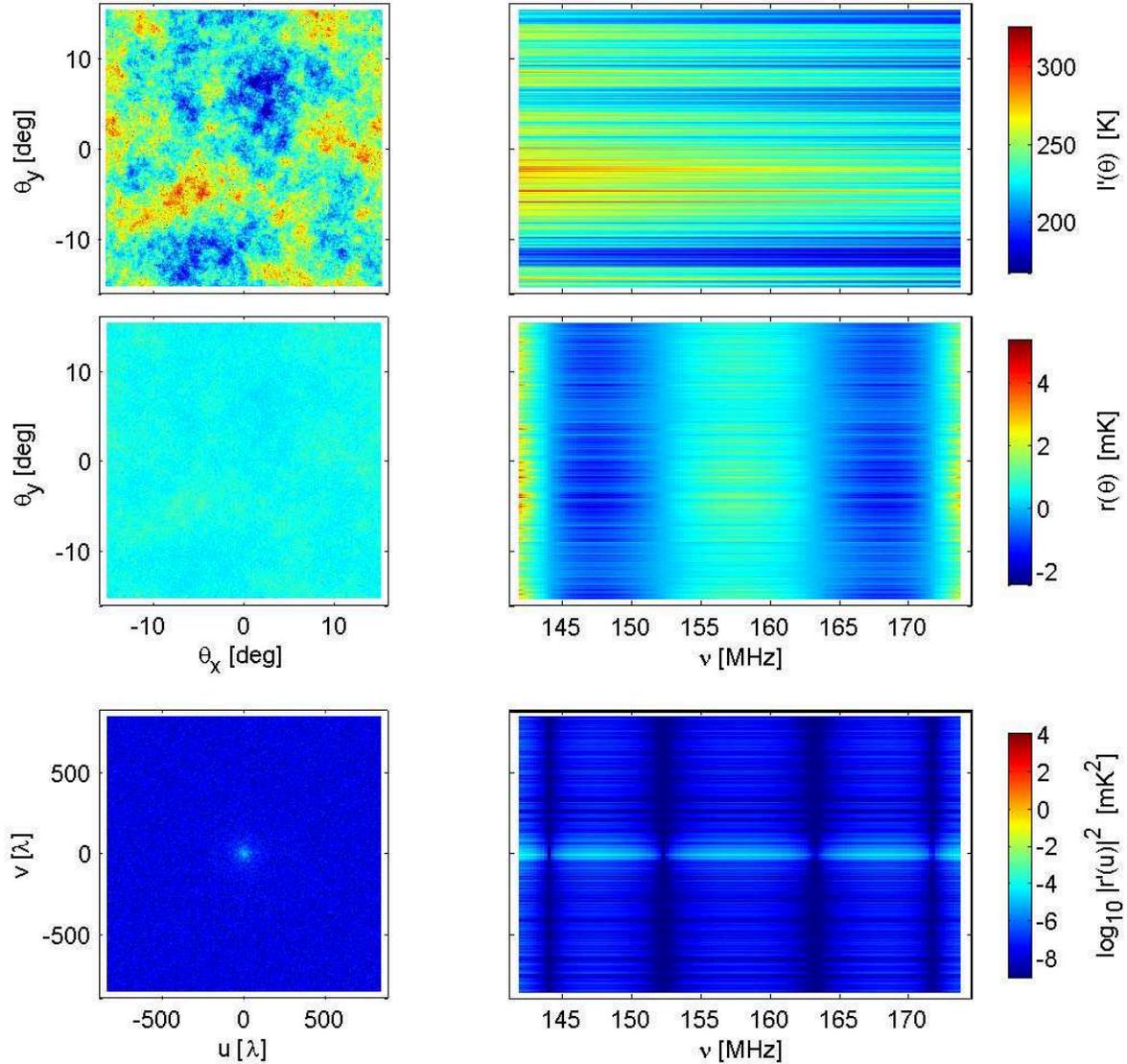}
\caption{\label{f_foregrounds_sky_model} \footnotesize Polynomial foreground subtraction technique applied to the input foreground sky model \textit{excluding} the simulated instrumental response.  Thermal noise and the 21~cm signal have been excluded from this simulation, as well, in order to provide a best-case reference for the foreground subtraction technique.  The first row shows two cuts through the foreground sky model.  The left panel is an image plane for a spectral channel at $\nu=157$~MHz, thus the pixels in the plot are indexed by $I'(\theta_x, \theta_y)$. The right panel is a slice through frequency and one angular dimension, thus the pixels in the plot are indexed through $I'(\nu, \theta_y)$.  The smooth spectral properties of the foreground model are easily visible in the righthand plot.  The middle row displays the residuals, $r$, in the image domain after the $3^\text{rd}$-order polynomial has been fit and subtracted from the spectra along each line of sight in the data cube.  Again, the left panel is an image plane and the right panel is a slice through frequency and one angular dimension.  The bottom row displays the residuals in the Fourier domain. Although it appears in the middle row that the residuals after foreground subtraction will be comparable to the $\sim25$~mK redshifted 21~cm signal, after transforming the residuals back to the Fourier domain (bottom row), it is evident that the polynomial fit and subtraction removed the foreground contribution well for all but the largest angular scales.}
\end{figure*}

Before we consider the effects of the instrumental response, we first review the ability of the polynomial fit technique to remove the foreground contributions in the image domain representation of our foreground sky model alone (with no 21~cm signal). This corresponds to the principal scenarios addressed by previous efforts \citep{2004MNRAS.355.1053D, 2004ApJ...608..622Z, 2004NewAR..48.1039F, 2004ApJ...608..611G, 2005ApJ...625..575S, 2006ApJ...650..529W, 2006ApJ...653..815M, 2008MNRAS.391..383G}.  Figure~\ref{f_foregrounds_sky_model} illustrates the results of performing the subtraction on the foreground-only sky model with no instrumental response.  The top row of the Figure shows the input sky data cube.  The left panel is a map from a single frequency channel, whereas the right panel is a $\nu$-$\theta_y$ plane that slices along the frequency axis.  The second row of the Figure shows the residuals in the data cube following the polynomial subtraction (again the left panel is a map from a single frequency channel, whereas the right panel is a slice along the frequency axis).  Finally, the bottom row of the Figure shows the Fourier domain representation of the residuals in the data cube.  The left panel in this case is the $uv$-plane for a single frequency channel and the right panel is a $\nu$-$v$ plane slicing along the frequency direction.  The residuals can be seen in the second row to be of order $\sim1$~mK.  This represents the best-case scenario that can be achieved by the $3^\text{rd}$-order polynomial fit method to our foreground model.  In general, the amplitude of the residuals could be made arbitrarily low by increasing the order of the polynomial used in the fits, but only at the expense of extracting additional power from the more rapidly varying spectral fluctuations. Since, in actual observations, the polynomial fit will remove power from the redshifted 21~cm fluctuations as well as the foreground contributions, it is desirable to use the lowest order polynomial feasible because this will limit the effects on the redshifted 21~cm power spectrum to only the longest length scales in the spectral domain.  The bottom panel shows that the residual angular power is peaked at large scales (low $u$).  The dark vertical bands in the bottom-right panel are at the zero-crossings of the residual power, which tend to occur at roughly the same frequencies for all sight-lines.  As was concluded by the previous efforts in the literature, it appears that, in an ideal case at least, the polynomial subtraction technique is sufficient to remove the foreground contribution in the sky to the level required to detect the expected 21~cm signal during reionization.

Next, we include the instrumental response in the analysis. Now there is an additional degree of freedom. The weighting function, $\mathbf{U}$, that is applied to the simulated gridded measurements before transforming from the Fourier to the image domain (see Equation~\ref{eqn_dirty_map}) must be specified before the calculation can be performed.  As discussed in the last section, we model uniform weighting, natural weighting, and a smoothed approximation to the natural weighting found by fitting a double exponential profile to the distribution of baselines (see Figure~\ref{f_foregrounds_weightprofiles}). The results of the foreground subtraction technique performed on the simulated dirty sky maps for each of these weightings are presented in Figures~\ref{f_foregrounds_sky_inst_uniform} through~\ref{f_foregrounds_sky_inst_exponential}, respectively.  Figure 10 shows a direct comparison of the dirty map subtraction residuals for the same line of sight using the three different weighting schemes.

\subsection{Uniform Weighting}

\begin{figure*}
\centering
\includegraphics[width=38pc]{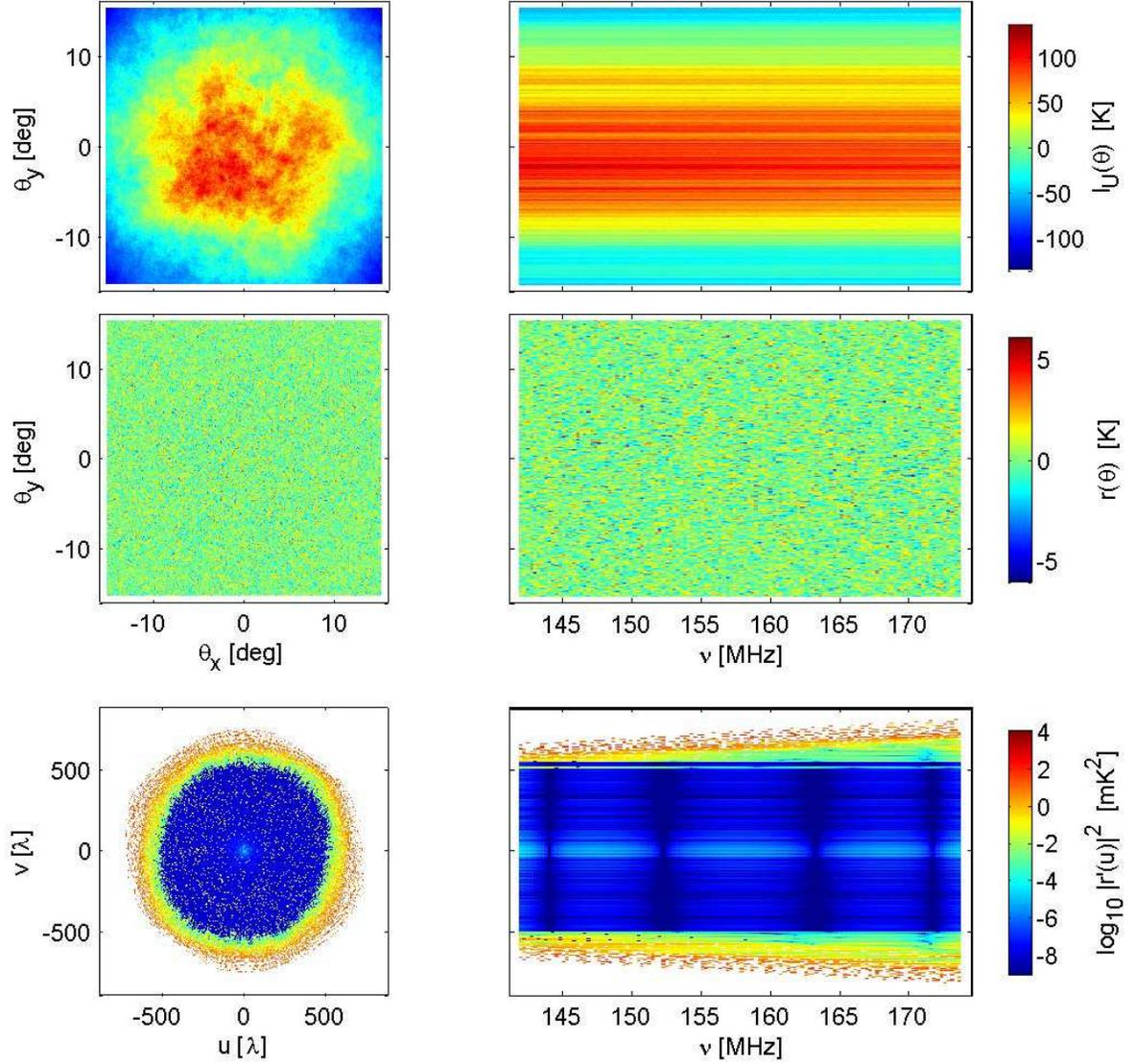}
\caption{\label{f_foregrounds_sky_inst_uniform} \footnotesize Polynomial foreground subtraction technique applied to the simulated dirty sky map generated with a uniform weighting of the visibility map.  The sky model used for the subtraction included only the foreground and thermal noise terms and \textit{excluded} the 21~cm signal.  The thermal noise contribution, however, has been artificially removed from these plots in order to better illustrate the foreground subtraction residuals.  In this case, the bottom row displays the unweighted residuals, $r'$, in the Fourier domain. Similar to the reference case in Figure~\ref{f_foregrounds_sky_model}, it appears in the middle row that the residuals after foreground subtraction will be much larger this time than the $\sim25$~mK redshifted 21~cm signal.  However, after transforming the residuals back to the Fourier domain and unweighting, we find that the polynomial fit and subtraction removed the foreground contribution well again for most baselines.  Aside from the short baselines that were poorly cleaned in even the instrument-free case, only  baselines longer than $\sim500\lambda$ are significantly contaminated when performing the subtraction in the dirty map generated with uniform weighting.}
\end{figure*}

\begin{figure*}
\centering
\includegraphics[width=38pc]{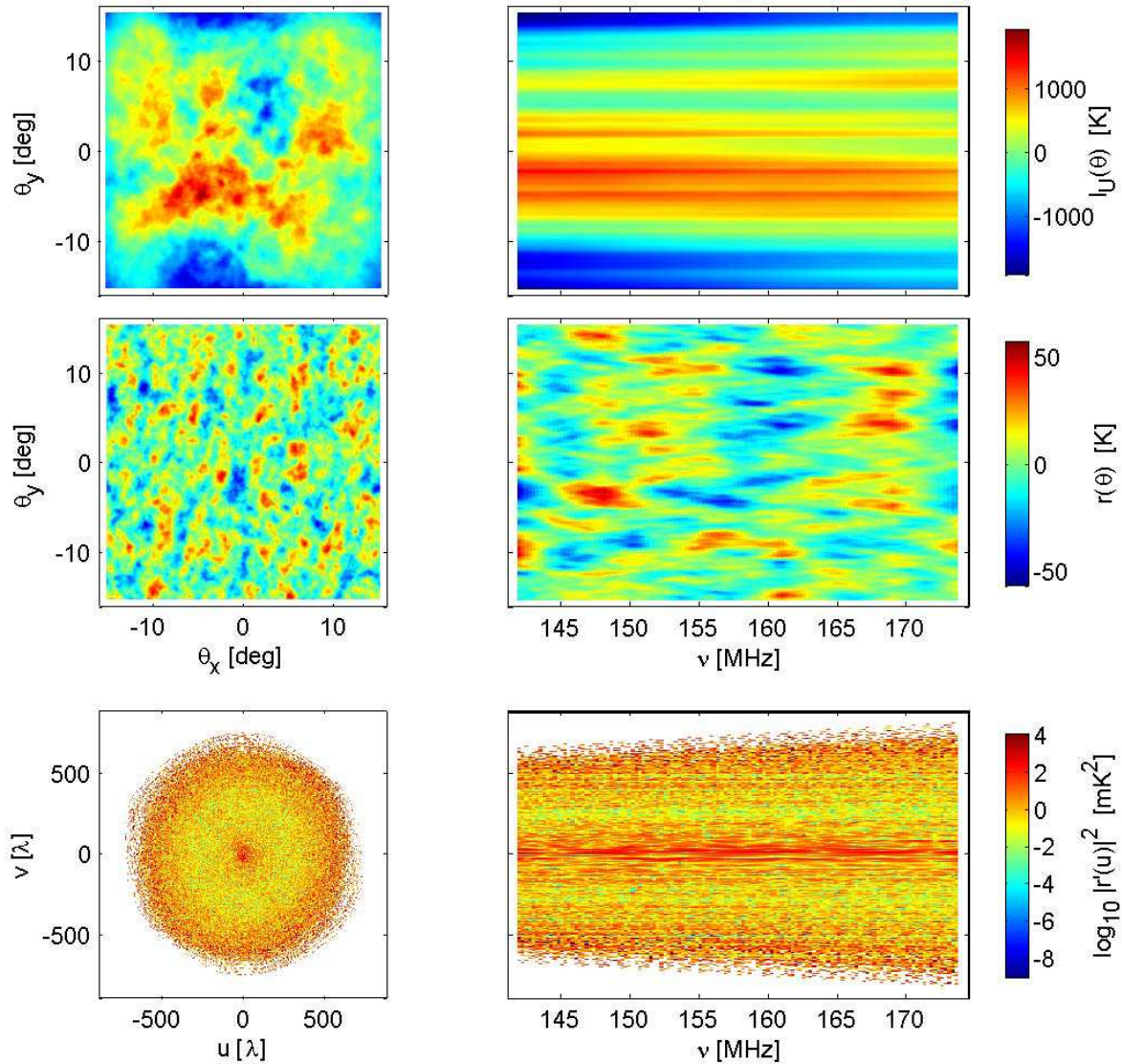}
\caption{\label{f_foregrounds_sky_inst_natural} Same as Figure~\ref{f_foregrounds_sky_inst_uniform}, but for the polynomial subtraction technique applied to the dirty sky map generated with a natural weighting of the visibility map.  No 21~cm signal was included in the analysis and the thermal noise has been artificially removed from the plots. As with the uniform weighting, the residuals, $r$, in the image domain appear much larger than for the best-case fits to the sky model alone.  In this case, however, even after transforming back to the Fourier domain and unweighting, the residuals remain much larger than for the other weighting methods.}
\end{figure*}

\begin{figure*}
\centering
\includegraphics[width=38pc]{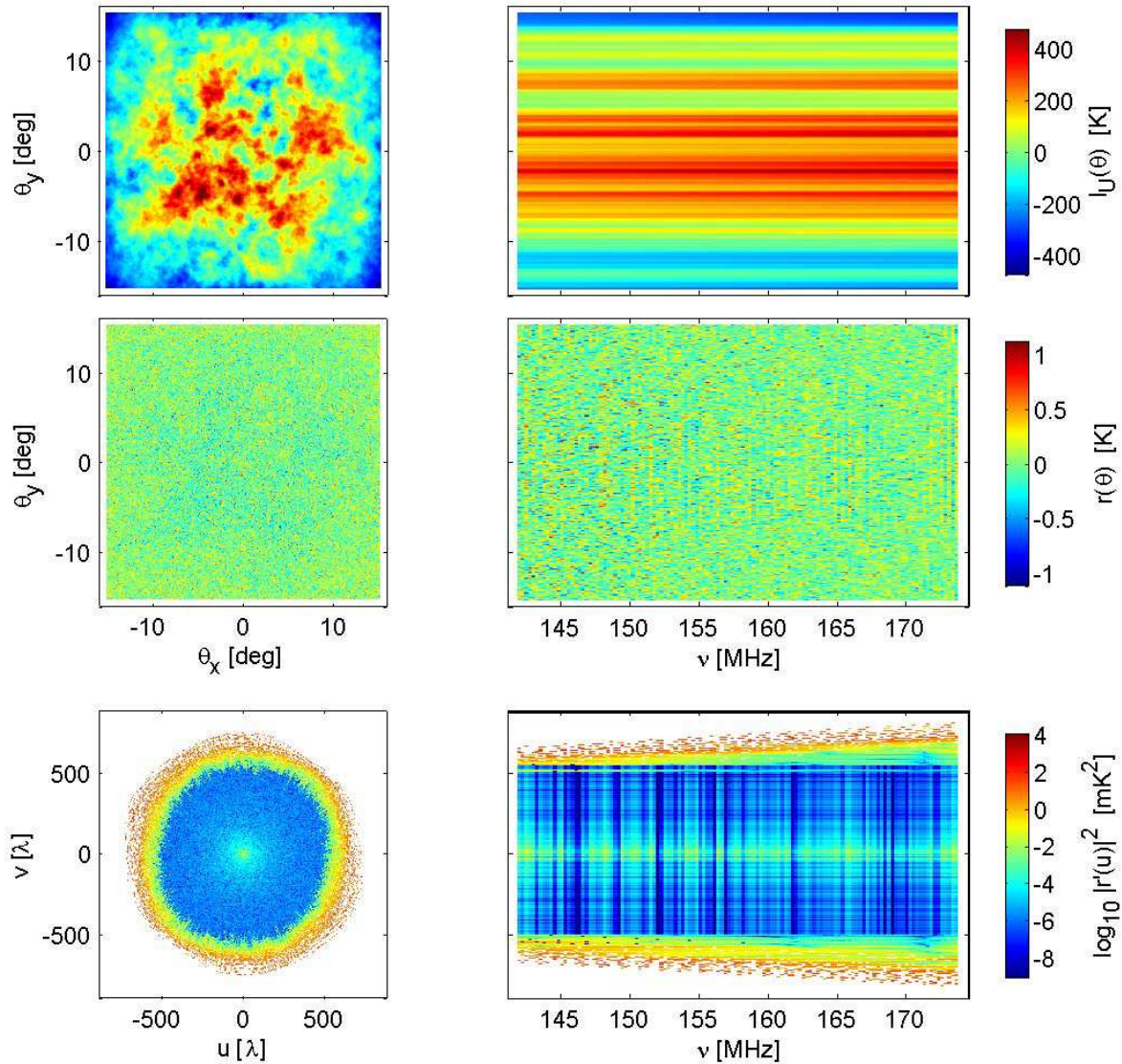}
\caption{\label{f_foregrounds_sky_inst_exponential} Same as Figure~\ref{f_foregrounds_sky_inst_uniform}, but for the polynomial subtraction technique applied to the dirty sky map generated with the smoothed approximation of the natural weighting of the visibility map.  No 21~cm signal was included in the analysis and the thermal noise has been artificially removed from the plots. For this case, the residuals, $r$, in the image domain are the lowest for the three dirty sky maps tested, but are still two orders or magnitude greater than the best-case subtraction for the sky model with no instrumental response. Transforming back to the Fourier domain and unweighting yields residuals that are nearly as low as the uniform weighting case.}
\end{figure*}

The first case is that of uniform weighting and is shown in Figure~\ref{f_foregrounds_sky_inst_uniform}. Here, the top row of the figure shows the dirty sky map, $I_D(\theta)$, produced by this weighting, the second row shows the results of the foreground subtraction in the image domain, $r(\theta)$, and the bottom row shows the results of the foreground subtraction after transforming back to the Fourier domain and removing the weighting, $r'(u)$.  In all cases, the thermal noise contribution has been artificially removed following the analysis to more clearly illustrate the foreground residuals.  As discussed in Section~\ref{sec_foregrounds_method}, this weighting gives the best effective angular resolution, but increases the uncertainty due to thermal noise in the dirty sky map.

At first glance, the results of the foreground subtraction in this case do not look especially promising since there are of order $\sim1$~K fluctuations in the residual map, $r(\theta)$, in the second row of Figure~\ref{f_foregrounds_sky_inst_uniform}.  However, after the residual map is transformed back to the Fourier domain, it becomes evident that the polynomial fit has actually done an excellent job of subtracting the foreground contamination from baselines within a radius of $u\lesssim500\lambda$, and only a poor job for baselines beyond this radius. The transition from the region where the foreground subtraction was successful to  the region where it failed is rapid and coincides with the radius where visibility measurements with the MWA become sufficiently sparse that there is no longer complete coverage.  In this outer region, a given $uv$-pixel is typically not sampled at all frequencies.  This creates variations at the corresponding angular scales between image maps at different frequencies and results in additional spectral structure in the image domain data cube.  Thus, on large angular scales, uniform weighting produces a dirty map with an effective PSF resembling a clean $\delta$-function yielding no corresponding sidelobe confusion since there is complete $uv$-coverage over all frequencies, whereas for small angular scales, the effective PSF is poorly behaved with large sidelobes.  This comes at the expense of elevated thermal noise in the dirty map due to the sub-optimal weighting of the $uv$-plane measurements when producing the map.

The bottom row of Figure~\ref{f_foregrounds_sky_inst_uniform} illustrates that it may not be necessary to selectively tailor the $uv$-plane coverage to be frequency independent for the MWA.  As long as the scales of interest are sampled completely for all frequencies, the sparsely sampled region does not interfere with the scales of interest following foreground subtraction.  Nevertheless, if the extra sidelobe noise in the dirty map due to the sparse coverage above $u\gtrsim500\lambda$ is not desired, then the long baselines could be completely discarded or selectively cut before producing the dirty map in order to yield a frequency-independent $uv$-coverage.  A variation of this approach was applied in \citet{2008MNRAS.389.1319J} for the planned LOFAR configuration.

\begin{figure}
\centering
\includegraphics[width=11pc,angle=-90,trim=1.82in 0 0 0,clip=true]{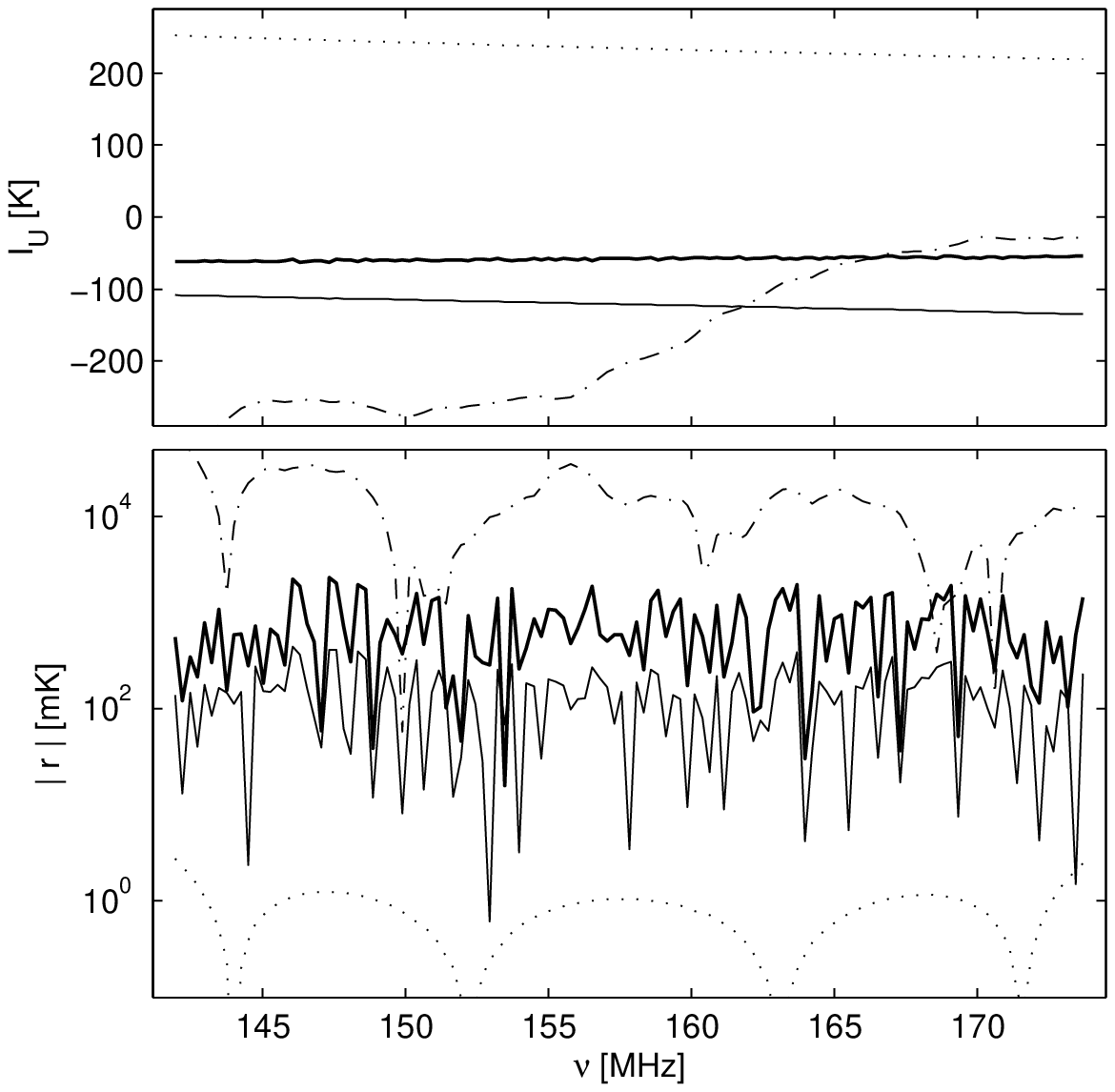}
\caption[Angular power spectra of foreground components]{
\label{f_foregrounds_subtractcompare} Polynomial subtraction residuals along the same line of sight in the dirty sky data cubes for the three weighting schemes as well as for the pure foreground sky model with no instrumental response simulated.  No 21~cm signal was included and thermal noise has been artificially removed.  The weighting schemes are uniform (thick solid line), natural (dash-dot line), exponential (thin solid), and the foreground model alone (dotted line). }
\end{figure}

\subsection{Natural Weighting}

The foreground subtraction results for the natural weighting case are shown in Figure~\ref{f_foregrounds_sky_inst_natural}.  It is clear even in the second row of Figure~\ref{f_foregrounds_sky_inst_natural} that the residual fluctuations in the dirty map are on much larger angular and spectral scales and have larger amplitudes than for the uniform weighting case, with $\sim10$~K fluctuations that are an order of magnitude greater than the expected 21~cm signal.  And in this case, the residuals persist at all scales after returning the Fourier domain and plotting $r'(u)$ in the third row of the figure. The best explanation for the poor subtraction obtained with the natural weighting can be found in the profiles of the weighting functions shown in Figure~\ref{f_foregrounds_weightprofiles}. The fine, hash-like fluctuations that are visible in the profile for the natural weighting due to the discrete antenna positions in the array are responsible for introducing ripples and fluctuations into the dirty sky cube generated with this weighting. So, although natural weighting would result in the least thermal uncertainty in the dirty sky map, it does so at the expense of coupling the variations in the density of visibility measurements in the $uv$-plane to the dirty sky map.  This is exactly what we would like to avoid in the sky map used for the foreground subtraction.

We can estimate the importance of structure in the $uv$ coverage at a given length scale. In general, the synthesized array beams at each frequency measured by an interferometer will be identical, but proportional in overall size to the inverse of the frequency, since the distribution of visibility measurements is identical, but scales with the wavelength in the $uv$-plane.  The coherence length of fluctuations in visibility coverage in the $uv$-plane is mapped to the spectral domain by this frequency-dependent scaling. The relationship between the coherence length in the $uv$-plane, $l_u$, and the coherence length in the frequency domain, $l_\nu$, is given by
\begin{equation}
\frac{l_u}{|u|} = \frac{l_\nu}{\nu_0}
\end{equation}
where $|u|$ is the radius of interest in the $uv$-plane at frequency $\nu_0$. From this expression, it is easy to see that modest features in the coverage of the visibility measurements near the origin of the $uv$-plane do not introduce significant fluctuations into the spectral domain because $l_\nu\rightarrow\infty$ as $|u|\rightarrow0$, whereas variations in the coverage or weighting at large radii in the $uv$-plane will have much shorter coherence lengths in the spectral domain because $l_\nu\rightarrow0$ as $|u|\rightarrow\infty$.  In order to produce fluctuations in the spectral domain that have sufficiently long periods that they can be fit by a low-order polynomial across the observed bandwidth requires $l_\nu\gtrsim B$, where $B$ is the bandwidth. For the MWA, with $\nu_0\approx150$~MHz and $B=8$ to 32~MHz, this condition requires that
\begin{equation}
\label{eqn_coherence_condition}
 l_u \gtrsim |u| \frac{B}{\nu_0} \approx \frac{|u|}{10}
\end{equation}
in order to prevent fluctuations in the spatial coverage from impacting the polynomial fits used in the foreground subtraction. Referring to Figure~\ref{f_foregrounds_weightprofiles}, this condition is clearly not met for the natural weighting function with its fine-scale features.

\subsection{Smoothed Natural Weighting}

We have seen so far that the uniform weighting allows foreground subtraction to succeed well below the required level, whereas the natural weighting does not.  However, the uniform weighting succeeds at the expense of increasing the thermal noise in the derived dirty map by nearly an order of magnitude compared to the natural weighting.  Since the natural weighting fails due to the fine-scale structure in the baseline distribution of the array, we explored smoothing the natural weighting function by approximating the distribution of baselines with a double exponential profile given by (see Figure~\ref{f_foregrounds_weightprofiles})
\begin{equation}
N(u) \sim e^{-|u| / l_1} + e^{-|u| / l_2}
\end{equation}
where $\l_1 = 90 \lambda$ and $\l_2 = 190 \lambda$, and the amplitude is set to match the true baseline distribution.  The function is cored for $|u| \leq 6\lambda$, approaching $0$ linearly, in order to prevent over-weighting these scales compared to the true natural weighting.  This weighting function retains the emphasis on the short baselines inherent in the natural weighting case, but without introducing the fine-scale variations in the density of the visibility measurements into the weighting function.  In our simulation, it resulted in only a 7\% increase in thermal noise in the dirty map compared to the true natural weighting.  The results of the foreground subtraction for this case are shown in Figure~\ref{f_foregrounds_sky_inst_exponential}.  This weighting function produces the best results in $r(\theta)$ and nearly matches the uniform weighting case in $r'(u)$.

\subsection{Review}

We pause for a moment to review the results of the analysis so far.  At this stage, we can conclude that subtraction of confusion-level foreground contamination in dirty sky maps produced by the MWA is possible using the simple ``per pixel'' polynomial subtraction technique.  Uniform weighting of the gridded visibility measurements during the dirty map generation process produces the best results of the three cases we tested.  If maximizing the S/N in the derived dirty map while preserving the purity of the spectral domain from mode-mixing contamination is desired, then we have shown that a smooth approximation to the natural weighting function yields similar results to the uniform weighting, but without significantly increasing the thermal noise in the dirty map above the best-case natural weighting level.  With additional studies, it seems likely that the smoothed variation of the natural weighting could be made to match the performance of the uniform weighting.

It may seem counter-intuitive that the success of the polynomial subtraction algorithm should be largely independent of the level of thermal noise in the input dirty map, so we note that this appearance is due, in part, to the fact that we have artificially removed thermal noise from the residuals plotted in Figures~\ref{f_foregrounds_sky_inst_uniform} through \ref{f_foregrounds_sky_inst_exponential} in order to facilitate an unhindered comparison of the different weighting schemes.  The derived foreground models do include the contributions of thermal noise, as do the resulting residuals.  However, our interest is not in the uncertainty of the foreground model, $d(\mathbf{\theta})$, but rather the uncertainty in the recovered 21~cm power spectrum, estimated from $r'(\mathbf{u})$.  We have thus far postponed our book-keeping of the thermal uncertainty so as not to confuse to the two.  In the remainder of this paper, we address this issue of thermal noise in the measurements and the ability of the MWA to detect the redshifted 21~cm signal after applying the foreground subtraction steps discussed above.  Since we have seen that the uniform weighting case is sufficient for foreground subtraction, we will use only uniform weighting in the analysis below for clarity of presentation.

\section{Recovering the EoR Signal}
\label{s_recovering}

After foreground subtraction, the $\sim10^8$ individual measurements comprising an MWA data cube contain largely thermal noise power and only a very small contribution from the redshifted 21~cm signal (and foreground residuals). To detect the EoR signal we need to average over all of these measurements to obtain 10--20 estimators with sufficient S/N to detect and characterize the reionization signal. The foundation for a statistical EoR power spectrum measurement of the brightness temperature fluctuations in low-frequency, wide-field radio observations has been developed in the literature by \citet{2004ApJ...615....7M} and \citet{2004ApJ...608..622Z}.  These efforts are built on the similar approach employed for interferometric measurements of CMB anisotropies \citep{1999ApJ...514...12W, 2002MNRAS.334..569H, 2003ApJ...591..575M}. The primary approach is to convert the three dimensional measurement cube to a one dimensional power spectrum.

Following foreground subtraction, the full 32~MHz data cube will be divided into 8~MHz sub-bands, each of which will be reduced into a power spectrum independently.  This approach is planned for the MWA analysis pipeline because the redshifted 21~cm signal is expected to evolve considerably over the $\Delta z\approx2$ redshift range spanned by the full 32~MHz observation.  By treating four sub-bands independently, any evolution of the signal can be largely neglected in the interpretation of the results since each sub-band will span only $\Delta z\approx0.5$.  The effect that this simplification has on the power spectra measurements is to eliminate measurements of the largest spatial scales, corresponding to the lowest $k_\|$ modes, along the line-of-sight direction.  Since these modes contain most of the foreground power, the are not expected to have yielded usable information about the 21~cm signal anyway.

Each residual data sub-cube, $r'(\{\u,\nu\})$, is transformed into three-dimensional wavenumber by applying a Fourier transform along the frequency-axis, $\nu\rightarrow\eta$, and then applying a coordinate transformation into cosmological coordinates $\k$:
\begin{equation}
r'(\k) = \mathbf{J}(\k,\{\u,\eta\}) \mathbf{F}(\{\u,\eta\},\{\u,\nu\}) r'(\{\u,\nu\}) \label{eqn_rk}
\end{equation}
where $\mathbf{J}(\k,\{\u,\eta\})$ is given by the Jacobian of the coordinate transformation from $\u,\eta$ (in units of $\lambda$ and Hz$^{-1}$) to $\k$ (in units of cMpc$^{-1}$, see \citet{2004ApJ...615....7M} for details). The residual $r'(\k)$ can be thought of as the three-dimensional Fourier transform of the foreground subtracted image cube as expressed in comoving Mpc.

Due to the isotropy of space, the power spectrum is approximately spherically symmetric in $\k$ coordinates. We can thus square $r'(\k)$ and average the result in spherical annuli
\begin{equation}
P(k) = \left < \left | r'(\k) \right |^2 \right >_{|\k|=k} \label{eqn_pk}
\end{equation}
to form the one dimensional power spectrum \citep{2004ApJ...615....7M}, or the more common dimensionless power spectrum given by $\Delta^2 = k^3 P(k) / (2 \pi^2)$.  However, it is useful to break the averaging from the three dimensional $\k$-space to the one dimensional $k$-space into two steps since both the foregrounds and a full treatment of the predicted redshifted 21 cm signal have \textit{aspherical} structure in the Fourier domain. Following the approach of \citet{2006ApJ...653..815M}, we first average over the angular sky direction to obtain $P(\kperp$, $\kpar)$. This is conceptually similar to averaging over the $m$ values and keeping the $\ell$ values in a CMB analysis, except we still have the line-of-sight dimension.  In practice, the 2D and 1D power spectra are determined following the maximum likelihood formalism that reduces to weighting the individual measurements by the inverse of the per-cell noise variance in the limit of no covariance terms between the cells.  For the 2D power spectrum this yields:
\begin{equation}
P\left(k_\bot, k_\|\right) = \frac{\sum_{|\mathbf{\kperp}|=k_\bot} \mathbf{U_n}(\k, \k)\left | r'(\k) \right |^2 }{\sum_{|\mathbf{\kperp}|=k_\bot} U_n(\k)} \label{eqn_pk2}
\end{equation}
and for the one dimensional power spectrum:
\begin{equation}
P(k) = \frac{\sum_{|\k|=k} \mathbf{U_n}(\k,\k)\left | r'(\k) \right |^2 }{\sum_{|\k|=k} U_n(\k)}, \label{eqn_pk1}
\end{equation}
where ${\mathbf{U_n}}$ is related to the natural weighting function used to generate dirty maps from the gridded visibility measurements, but with spectral structure converted into spectral covariance terms by the Fourier transform along the frequency axis.  Neglecting the effects of RFI and RRL excision, the $k_\|$ covariance is highly peaked for the MWA due to the dense $uv$-coverage and we make the simplifying approximation in Equations~\ref{eqn_pk2} and \ref{eqn_pk1} to treat it as a delta function in this analysis so that the covariance matrix for the $\k$-space Fourier data cube is taken to remain diagonal and $\mathbf{U_n})$ can be treated as a multiplicate operator independent of $k_\|$.

Theoretical models indicate that the true redshifted 21~cm signal will be composed of contributions (due to velocity field and other astrophysical effects) from modes that are modulated according to powers of $\mu\equiv\hat{k}\cdot\hat{n}$, where $\mu$ is the cosine of the angle between the line of sight and the wavevector \citep{1987MNRAS.227....1K, 2005ApJ...624L..65B, 2006ApJ...653..815M}.  Thus,
\begin{equation}
P_{21}\left(\vec{k}\right) = P_{\mu^0}(k) + \mu^2 P_{\mu^2}(k) + \mu^4 P_{\mu^4}(k) + \ldots, \label{eqn_mu}
\end{equation}
and can also be fully described by $P_{21}(k_\bot, k_\|)$.  The simple model of the 21~cm signal we are using in this paper, however, contains only a single $P_{\mu^0}$ contribution and the MWA will have insufficient sensitivity initially to detect the higher-order aspherical effects independently.

Even in reduced power spectra, the 21~cm signal is weaker than the total thermal noise power and additional steps must be taken to isolate it.  There are two possible approaches for extracting the 21~cm contribution from the noise-dominated measurement: 1) calculate the thermal noise contribution to the auto-correlation generated power spectrum and subtract it from the final result, or 2) generate the power spectrum by dividing the observation into two epochs of equal duration and then cross-correlating the data cubes from the two epochs (M. Tegmark, private communication).  This second approach preserves the persistent 21~cm signal and eliminates the thermal noise power (which will be independent between the two observing epochs and, therefore, average to zero during the cross-correlation), leaving only the thermal uncertainty.  It has the advantage that the system temperature does not need to be known independently to high precision in order to recover the 21~cm signal (as long as the thermal noise properties of the instrument do not vary between epochs).  We have chosen to follow this second approach and calculate the 1D and 2D power spectra as cross-power spectra by replacing the $|r'(\k)|^2$ terms in Eqns~\ref{eqn_pk}--\ref{eqn_pk1} with $(r_1'^{\ast} r_2' + r_2'^{\ast} r_1')/2$ such that Eqn~\ref{eqn_pk} becomes:
\begin{equation}
P(k) = \left < \frac{r_1'^{\ast} r_2' + r_2'^{\ast} r_1'}{2} \right >_k
\end{equation}
where $r'_1$ and $r'_2$ are the residuals for the two epochs of the simulated observation.

\begin{figure*}
\centering
\includegraphics[width=40pc, angle=-90]{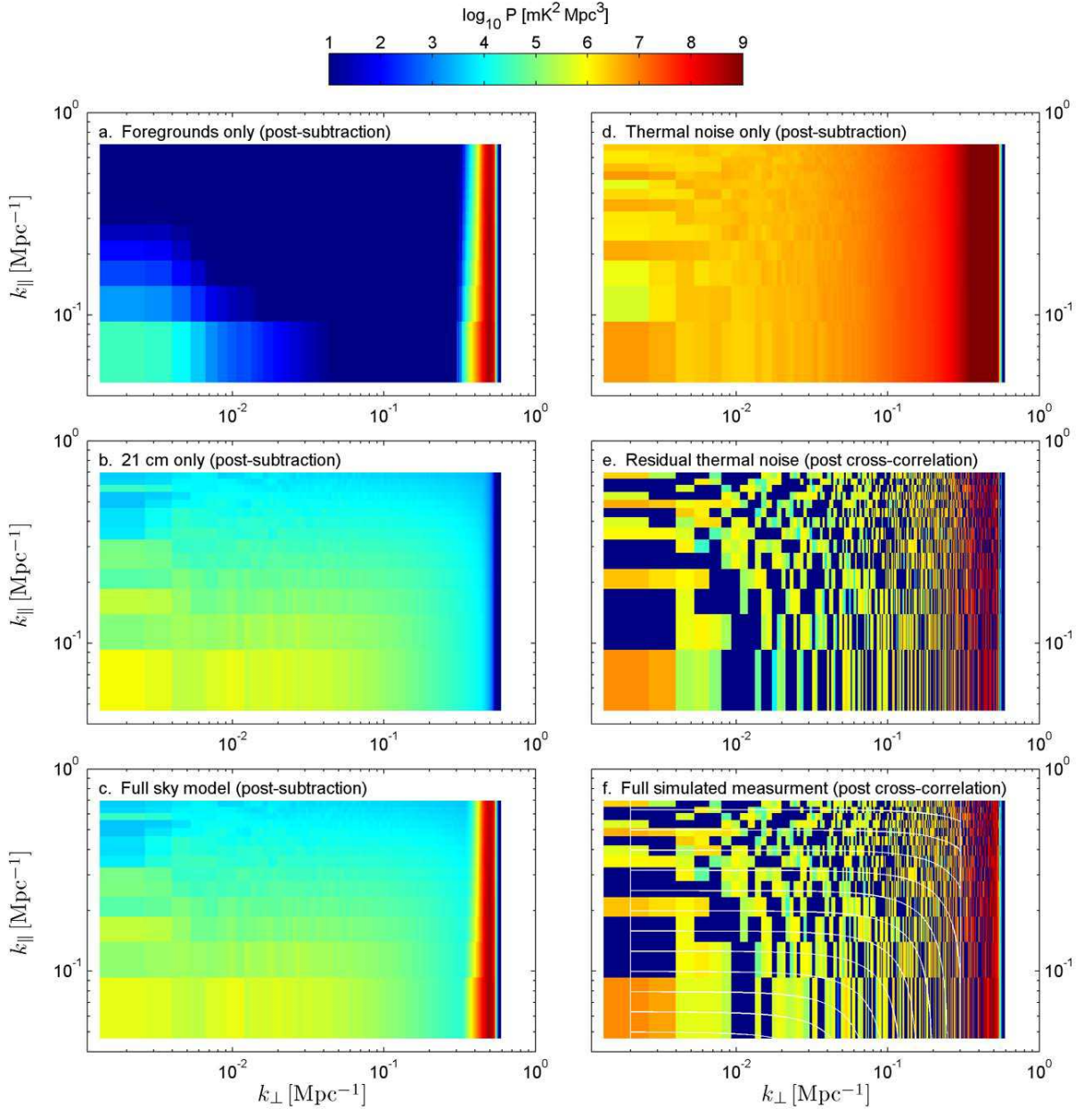}
\caption{ \label{f_foregrounds_2d_1} 2D power spectra of post-subtraction residuals from dirty maps generated with uniform weighting.  For the left column of panels, the thermal noise has been artificially removed.  It is evident that the foreground-only residuals following subtraction are much lower than the 21~cm signal in the range $k_\bot \lesssim 3\times10^{-1}$, beyond which corresponds to the outer anulus of poor foreground subtraction in the bottom panel of Figure \ref{f_foregrounds_sky_inst_uniform}.  Panel \textit{c} confirms that the 21~cm signal dominates the recovered power from the full sky model since it appears nearly identical to panel \textit{b}.  The white arcs in panel \textit{f} illustrate the spherical shells of constant-$k$ that are used for the 1D power spectrum in Figure~\ref{f_foregrounds_ps_sim}.}
\end{figure*}

The panels in Figure~\ref{f_foregrounds_2d_1} illustrate the post-subtraction residuals in the 2D power spectrum of the full simulated instrumental response.  The left column of panels show the post-subtraction power spectra of the individual and combined sky model components with: a) the foreground-only contribution to the sky model, b) the redshifted 21~cm-only contribution, and c) the full combined sky model.  Thermal noise has been artificially removed from these three panels to highlight the structure of the astrophysical sky.  It is clear that the residual power in the post-subtraction, foreground-only contribution in panel (a) is well below the 21~cm-only power in panel (b).  Panel (c) confirms that the foreground subtraction recovers the 21~cm signal since the subtraction results on the full sky model are nearly identical to the 21~cm-only results.  The residual foreground contamination in panel (a) does not have the same symmetry as the redshifted 21~cm signal, and tends to have functional forms given by $P(\k) = P(k_\bot)P(k_\|)$, as opposed to the spherical and $\mu$-modulated 21~cm symmetries. \citet{2006ApJ...648..767M} have discussed using these differences in shape to further separate the 21~cm power spectrum from the foregrounds, although we do not pursue this step in the present analysis.

\begin{figure}
\centering
\includegraphics[width=15pc, angle=-90]{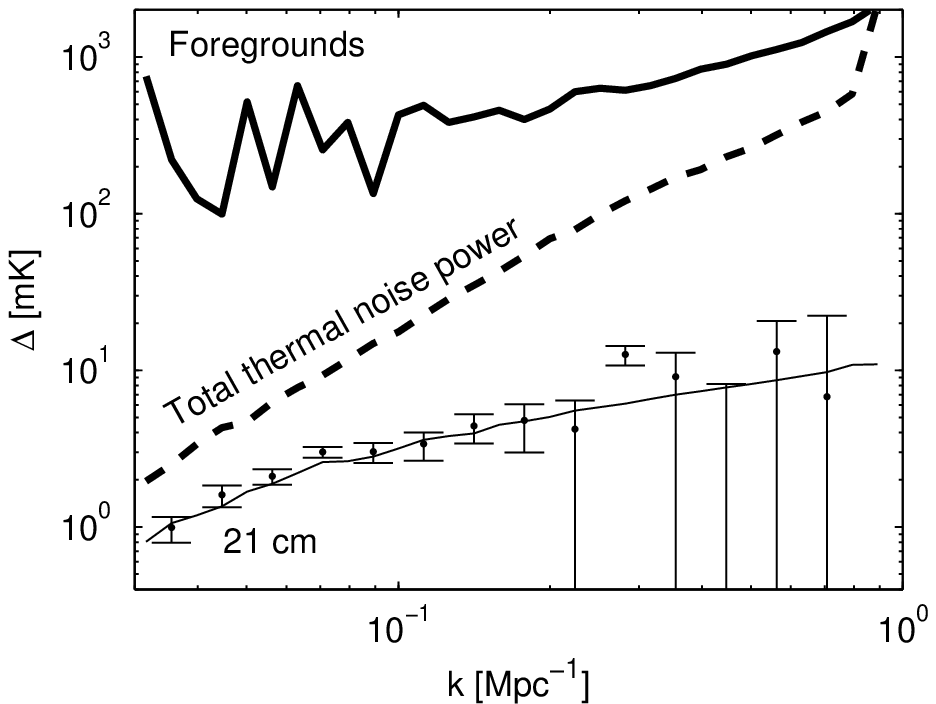}
\caption{ \label{f_foregrounds_ps_sim} Simulated measurement of the 1D spherically binned redshifted 21~cm power spectrum by the MWA in a single 8~MHz sub-band with 360~hours of integration.  The thin solid line is the input 21~cm signal.  The error bars give the thermal uncertainty on the recovered measurement and are $1\sigma$.  Measurements below $k \lesssim 10^{-1}$ Mpc$^{-1}$ are highly coupled due to the limited depth of the observed comoving volume, as is evident from the constant-$k$ contours in the bottom-right panel of Figure~\ref{f_foregrounds_2d_1}. The thick solid line shows the total power (before subtraction) in the simulated sky data cube including thermal noise and foregrounds, and the thick dash line shows the total power of the thermal noise alone.}
\end{figure}

The right column of panels in Figure~\ref{f_foregrounds_2d_1} illustrates the thermal noise properties and the removal of the mean thermal noise power from the data cube by the cross-correlation.  Panel (d) shows the total thermal noise in the simulated power spectrum after 360~hours of observation with the MWA.  Despite the long integration, the total thermal noise still dominates the desired 21~cm signal by over an order of magnitude.  After dividing the simulated observation into two epochs and calculating the cross-correlation, however, the mean thermal noise power is removed from the final power spectrum, leaving only the residual thermal uncertainty power shown isolated in panel (e) and with the full simulated measurement including the sky signal in panel (f).  The MWA will not have enough sensitivity within its first year of observations to produce a high S/N 2D power spectrum so the 1D power spectrum will be the primary output product.  The white curves in panel (f) indicate the spherical shells of constant-$k$ that are used to calculate the 1D power spectrum in Figure~\ref{f_foregrounds_ps_sim}.  In addition to the recovered 21~cm power spectrum, Figure~\ref{f_foregrounds_ps_sim} also illustrates the total sky power and total thermal noise power in the 1D power spectrum before subtraction and cross-correlation.

We note that, during a year-long season of observing on the primary target window, the MWA should produce four power spectrum measurements, each comparable to the simulated results in Figures~\ref{f_foregrounds_2d_1}f and \ref{f_foregrounds_ps_sim}, spanning the full redshift range $7.1 < z < 9$ covered by the 32~MHz instrumental bandwidth in blocks of $\Delta z\approx0.5$, corresponding to the individual 8~MHz sub-bands of the divided data cube.  In addition, a secondary target field should yield an independent set of observations producing four additional similar power spectra, for a total of eight unique power spectra documenting the 21~cm emission during the reionization epoch.

\subsection{Characterization of Subtraction Effects}
\label{sec_foregrounds_bias}

The foreground subtraction process introduces errors into the uncorrected derived estimate of the redshifted 21~cm power spectrum. The two principal sources of possible error are: 1) the unavoidable removal of some of the large-scale power from the redshifted 21~cm signal by the polynomial fitting and subtraction algorithm, and 2) the contamination of the derived power spectrum by any residual foreground power resulting from an imperfect subtraction.  These effects must be quantified through modeling to accurately interpret real measurements.  As an initial step, we introduce a subtraction characterization factor, $f_s$, that captures the ratio of the recovered power spectrum to the true power spectrum such that
\begin{equation}
f_s(\k) = P_{21}(\k) / P'_{21}(\k), \label{eqn_bias}
\end{equation}
where $P'_{21}$ is the true input 21~cm power spectrum, and $P_{21}$ is the recovered output power spectrum following the simulated measurement and foreground subtraction process.

\begin{figure}
\centering
\includegraphics[width=15pc, angle=-90]{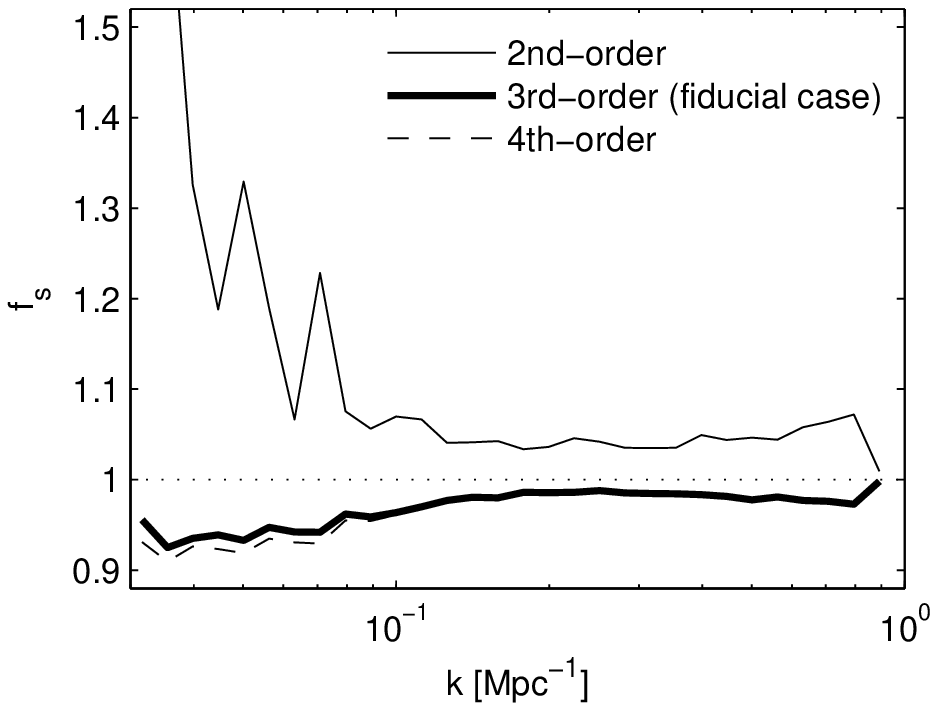}
\caption{ \label{f_foregrounds_1d_bias1} Subtraction characterization factor, $f_s$, for the 1D binned power spectrum.  The characterization factor is shown for three different levels of polynomial subtraction.  As expected, the largest error is for large spatial modes (low $k$).  Over most scales, however, the correction for our fiducial 3$^\textrm{rd}$-order polynomial subtraction is only $\sim1$\%.}
\end{figure}

Figure \ref{f_foregrounds_1d_bias1} plots the net characterization factor calculated for our fiducial model.  The bias due to performing the polynomial fit and subtraction in the presence of foregrounds reduces the overall power in the recovered power spectrum at all scales (such that $f_s(k)<1$), although this effect is minor---less than $10$\% at any scale, and of order only $\sim1$~\% for $k\gtrsim10^{-1}$~Mpc$^{-3}$.  This suggests that the polynomial subtraction process is removing more power from the 21~cm signal than any residual foreground power following subtraction is adding to the recovered estimate.  This conclusion is confirmed when we calculate the characterization factor for the recovered signal without any foregrounds in the sky model and find less than a 0.1\% difference between this 21~cm-only case and the result for the full sky model shown in the Figure.

As a counter point, we include in Figure~\ref{f_foregrounds_1d_bias1} a separate curve (thin solid line) for the characterization factor calculated for the subtraction process on the full sky model using only a 2$^\text{nd}$-order polynomial.  In this case, residual foreground power resulting from the poorer fit by the 2$^\text{nd}$-order polynomial contributes more to the recovered 21~cm power spectrum than the polynomial fit itself removes and yields $f_s>1$ over the entire spectrum.  This is particularly evident for $k<10^{-1}$~Mpc$^{-3}$, where it is clear that the lower-order polynomial failed to remove much of the foreground power.  Finally, we note that increasing the polynomial to 4$^\textrm{th}$-order has a minor effect on the recovered 21~cm signal (shown as the dashed line in Figure~\ref{f_foregrounds_1d_bias1}) that is evident only at large scales.  This suggests that there is additional margin in the polynomial subtraction approach to handle more complicated foregrounds than currently anticipated without substantially degrading the recovered 21~cm signal.

In principle, the characterization factor is dependent not only on the foreground subtraction algorithm, but also on the specific shapes of the redshifted 21~cm signal and the foregrounds.  Since we have assumed a simple 21~cm model with structure based only on a Gaussian random field of matter density perturbations as well as a fairly general model of the foreground contribution, the $f_s$ derived from this analysis should be taken only as a guide.

\section{Discussion}

In the preceding sections, we have shown that the effects of the frequency-dependent instrumental response of the MWA do not hinder the simple polynomial-fit diffuse foreground subtraction scheme to recover the redshifted 21~cm signal.  This is particularly encouraging because the fit has been performed in dirty sky maps, which are very sensitive to instrumental effects.  Thus, we have established, in some sense, a worst-case scenario that appears manageable.

Dense $uv$-coverage is a new advancement for radio arrays.  In the traditional paradigm, the $uv$-plane is almost always sparsely sampled (as with the VLA and other existing arrays), and there are essentially no regions where the coverage can be said to be ``complete'', as is the case for the MWA within $u\lesssim500\lambda$.  The reason that this is generally not a problem for telescopes like the VLA is that, in most cases, these instruments are not attempting to make observations at or below the confusion limit imposed by faint extragalactic continuum sources and diffuse Galactic structure. So, although the VLA has sparse coverage in the $uv$-plane, the sky that it is observing is also essentially sparse and deconvolution algorithms are able to reconstruct a reasonable model of the point sources in the field of view.

It is useful to construct an analytic approximation for the simulations discussed in the previous sections in order to foster a more intuitive understanding of the interaction between the array properties and the success of the foreground subtraction. \citet{1974ApJ...188..279C}, \citet{vla-sci-146}, and \citet{2003A&A...409..787A} provide a foundation for such a model, and Sault and Wayth [2006, 2007, MWA project documents] have adapted the results of these efforts for analyzing confusion noise in measurements of the MWA.

As we outlined in Section~\ref{sec_foregrounds_method}, a dirty sky map produced by an interferometer is, neglecting calibration errors, the result of multiplying the true sky by the primary antenna tile beam and then convolving by the synthesized array beam.  If the sky consists only of the faint extragalactic continuum sources with Poisson statistics, then the variance in the intensity of the dirty sky map, $\sigma^2_D$, will be related to the variance in the true sky by
\begin{equation}
\label{eqn_confusion_analytic_full} \sigma^2_D(\theta_x, \theta_y) =
\sigma^2_T \int \int B^2(\theta_x-\theta_x', \theta_y -
\theta_y')~P^2(\theta_x, \theta_y) ~ d\theta_x ~ d\theta_y
\end{equation}
where $\sigma^2_T$ is the variance in the intensity of the true sky, $B$ describes the response of the synthesized array beam as a function of angle, and $P$ describes the primary antenna tile beam. This equation can be simplified by assigning simple models for the response profiles of the synthesized array beam and the primary antenna tile beam. We take the beams to be described by tophat functions, such that the response is defined to be one within a region of diameter $\Theta_B$ (for the synthesized array beam; $\Theta_P$, for the primary antenna tile beam). Outside this region, the response is taken to be $B_{rms}\ll 1$ for the synthesized beam, and zero for the primary antenna tile beam. Thus, we include an allowance for the side lobes of the synthesized beam, but not for the primary beam. With these simplifications, the integrals in Equation~\ref{eqn_confusion_analytic_full} can be performed and the
expression reduces to
\begin{equation}
\sigma^2_D \approx \sigma^2_S \left(1 + B^2_{rms} \Omega_P
/\Omega_B\right )
\end{equation}
where $\sigma^2_S$ is the variance due to the faint sources in our model and accounts for the extra variance due to the weak angular clustering, $\Omega_B\approx\Theta^2_B$ is the solid angle of the synthesized array beam, and $\Omega_P\approx\Theta^2_P$ is the solid angle of the primary antenna tile beam.

Inspecting this simplified form of Equation~\ref{eqn_confusion_analytic_full}, we see that when the
synthesized array beam has no side lobes, the variance in the dirty sky map is equal to that of the true sky map (after gridding). This is equivalent to complete coverage of the $uv$-plane and corresponds generally to the cases considered by \citet{2004MNRAS.355.1053D, 2004ApJ...608..622Z, 2004NewAR..48.1039F, 2004ApJ...608..611G, 2005ApJ...625..575S, 2006ApJ...650..529W, 2006ApJ...653..815M, 2008MNRAS.391..383G}.  But when side lobes are included in the model of the synthesized array beam, the variance in the dirty map is increased above that of the true sky map.  For sparse coverage in the $uv$-plane, $B_{rms}^2$ will be given approximately by the inverse of the number of independent measurements in the $uv$-plane. In the limiting case of a single, very short integration, (such as a ``snapshot'' observation), the number of independent measurements is equal to the number of baselines. For the MWA, with its $\sim$125,000 instantaneous baselines, $B^2_{rms}\approx10^{-5}$ in the worst-case, and is even better (lower) for long integrations when earth-rotation synthesis increases the number of independent measurements.  Using $\Theta_B\approx5'$ and $\Theta_P\approx30^\circ$, we can estimate for the MWA that
\begin{equation}
B^2_{rms} \Omega_P /\Omega_B \approx 1.
\end{equation}
Thus, the variance due to the side lobes in the synthesized array beam is comparable in magnitude to the inherent variance in the intensity of the true sky.  A large field of view acts to oppose the advantages of dense coverage of the $uv$-plane in this instance by introducing more variance into the dirty sky map. In this regime, in order to produce an estimate of the sky without the effects of the side lobes of the synthesized array beam, the full inversion of $\mathbf{A}$ is required since it cannot be approximated as a sparse matrix.

Nevertheless, side lobe-induced noise in the dirty sky maps does not necessarily mean that the foreground subtraction
will be adversely affected. If the side lobe patterns were the same for each image plane in the data cube, then the spectrum along each pixel would simply receive a faint mirror contribution from the smooth spectra along all the other pixels.  It is the variation with frequency of the side lobe-induced noise that can cause trouble in the planned foreground subtraction technique.  As we found in Section~\ref{sec_foregrounds_subtraction}, however, the density of the MWA $uv$-coverage effectively eliminates the frequency-dependence of the sidelobe structure of the array beam over the large range of scales where the $uv$-plane is completely sampled at all frequencies.  Although it does not appear to be necessary for the MWA, less densely sampled arrays could control the spectral dependence of their array beam sidelobes by masking out regions of the $uv$-plane that are not sampled at all frequencies and only processing regions that do have coverage in each spectral channel (as considered in \citet{2008MNRAS.389.1319J} for the more sparsely sample LOFAR).

Despite the increased susceptibility of long baselines to introducing undesirable fluctuations into the spectral domain, it is important to point out that they serve an important purpose.  Long baselines are still expected to be necessary in order to peel away the bright sources in the field of view since the peeling process requires a precise knowledge of the position of each bright source, which is greatly improved with long baselines.

It has been shown previously that the ideal baseline distribution for minimizing thermal uncertainty in measurements of the redshifted 21~cm 1D power spectrum is extremely condensed \citep{2006ApJ...638...20B, 2007arXiv0711.4373L}.  In this work, we have further demonstrated that a compact array like the MWA also produces the dense $uv$-coverage necessary to efficiently subtract faint and diffuse foregrounds from dirty maps.

\acknowledgements

This work was supported by the Massachusetts Institute of Technology, School of Science, and by the NSF through grant AST-0457585. JDB is supported by NASA through Hubble Fellowship grant HF-01205.01-A awarded by the Space Telescope Science Institute, which is operated by the Association of Universities for Research in Astronomy, Inc., for NASA, under contract NAS 5-26555.


\end{document}